\documentstyle[multicol,eqsecnum,psfig,epsf,aps]{revtex}
\newcommand{\be}{\begin{equation}}
\newcommand{\ee}{\end{equation}}
\newcommand{\bea}{\begin{eqnarray}}
\newcommand{\eea}{\end{eqnarray}}
\newcommand{\ba}{\begin{array}}
\newcommand{\ea}{\end{array}}
\newcommand{\bi}{\begin{itemize}}
\newcommand{\ei}{\end{itemize}}
\newcommand{\eqnMHDubx}
{\bea 
\frac{\partial {\bf u}}{\partial t} +
{\bf (u \cdot \nabla) u} & = & 
- \nabla p + {\bf (b \cdot \nabla) b} + \nu \nabla^2 {\bf u} 
		\label{eqn:udot_x} \\
\frac{\partial {\bf b}}{\partial t} +
{\bf (u \cdot \nabla) b} & = & 
- {\bf (b \cdot \nabla) u} + \eta \nabla^2 {\bf b}
		\label{eqn:bdot_x} \\
\nabla \cdot {\bf u} & = & 0 \\
\nabla \cdot {\bf b} & = & 0  \eea}

\begin{document}
\def\slash#1{\setbox0=\hbox{$#1$}#1\hskip-\wd0\hbox to\wd0{\hss\sl/\/\hss}}

\title{Energy transfer in two-dimensional magnetohydrodynamic turbulence:
formalism and numerical results}
\author{Gaurav Dar, 
Mahendra K. Verma  \thanks{e-mail: mkv@iitk.ac.in} \\ 
Department of Physics \\ 
Indian Institute of Technology, 
Kanpur 208016, India \\
V. Eswaran  \\ 
Department of Mechanical Engineering \\ 
Indian Institute of Technology, 
Kanpur 208016, India}
\maketitle

\begin{abstract}

The basic entity of nonlinear interaction in Navier-Stokes and the
Magnetohydrodynamic (MHD) equations is a wavenumber triad ({\bf
k,p,q}) satisfying ${\bf k+p+q=0}$. The expression for the combined
energy transfer from two of these wavenumbers to the third wavenumber
is known. In this paper we introduce the idea of an effective energy
transfer between a pair of modes by the mediation of the third mode,
and find an expression for it. Then we apply this formalism to
compute the energy transfer in the quasi-steady-state of
two-dimensional MHD turbulence with large-scale kinetic forcing.  The
computation of energy fluxes and the energy transfer between different
wavenumber shells is done using the data generated by the pseudo-spectral
direct numerical simulation.  The picture of energy flux that emerges
is quite complex---there is a forward cascade of magnetic energy, an
inverse cascade of kinetic energy, a flux of energy from the kinetic
to the magnetic field, and a reverse flux which transfers the energy
back to the kinetic from the magnetic. The energy transfer between
different wavenumber shells is also complex---local and nonlocal
transfers often possess opposing features, i.e., energy transfer
between some wavenumber shells occurs from kinetic to magnetic, and
between other wavenumber shells this transfer is reversed. The net
transfer of energy is from kinetic to magnetic. The results obtained
from the studies of flux and shell-to-shell energy transfer are
consistent with each other.
\end{abstract}

\vspace{1.0cm}


\section{Introduction}
\label{sec:intro}

In fluid and MHD turbulence, eddies of various sizes interact amongst
themselves; energy is exchanged among them in this process. These
interactions arise due to the nonlinearity present in these systems.
The fundamental interactions involve wavenumber triads ({\bf k,p,q})
satisfying ${\bf k+p+q=0}$.  The {\it combined energy transfer}
computation to a mode from the other two modes of a triad has
generally been considered to be fundamental.  This formalism has
played an important role in furthering our understanding of
interactions in fluid turbulence.  In this paper we present a new
scheme to calculate the energy transfer rate between two modes in a
triad interaction in magnetohydrodynamic (MHD) turbulence---we call
it the {\it mode-to-mode transfer}. Using our scheme we calculate
cascade rates and energy transfer rates between two wavenumber shells.

The energy spectra and cascade rates are quantities of interest in the
statistical theory of MHD turbulence.  Recent numerical
\cite{MKV:MHDsimu,Bisk1:Kolm,Bisk2:Kolm} and theoretical
\cite{MKV:B0RG,MKV:MHD_PRE,MKV:MHDRG} work indicate that
the energy spectrum of total energy is proportional to $k^{-5/3}$,
similar to that found in fluid turbulence. Sridhar and Goldreich
\cite{Srid1} and Goldreich and Sridhar \cite{Srid2} studied
the energy spectrum in presence of strong magnetic field and 
found it to be proportional $k_{\perp}^{-5/3}$ ($k_{\perp}$
is the perpendicular component of the wavevector).  
Most of the work
on MHD turbulence focus on the energy spectrum of either total energy
or on the spectra of Els\"{a}sser variables (${\bf u \pm b}$, where
{\bf u} and {\bf b} are velocity and magnetic fields respectively).
In this paper we will focus on various energy cascade rates of MHD
turbulence.  The magnetic energy (denoted by ME) of a mode evolves due
to two nonlinear terms, ${\bf [b.(u. \nabla) b]}$ and ${\bf
[b.(b. \nabla) u]}$, of the MHD equations; the first term exchanges ME
between different scales, and the second term exchanges magnetic and
kinetic energy (denoted by KE) between different scales.  The 
KE similarly evolves due to two nonlinear terms ${\bf
[u.(u. \nabla) u]}$ and ${\bf [u.(b. \nabla) b]}$; the first one
exchanges KE between different scales and the second
exchanges energy between the magnetic and the velocity fields.

Pouquet {\it et al.}~\cite{Pouq:EDQNM} studied the energy transfer
between large and small scales using three-dimensional (3D) EDQNM
closure calculations. For nonhelical MHD they argued that the ME
cascades forward, i.e., from large scales to small scales.  In
presence of helicity, the large-scale magnetic field grows, which in
turn brings equipartition the small-scale kinetic and magnetic energy
by Alfv\'{e}n effect.  The residual helicity, the difference between
kinetic and magnetic helicity\footnote{Kinetic helicity= $\int {\bf
u.\omega} d^{3}x$, where ${\bf \omega = \nabla \times u}$.  Magnetic
current helicity = $\int {\bf b.j} d^{3}x$, where ${\bf j = \nabla
\times b}$} , induces growth of large-scale magnetic energy and
helicity.  Pouquet and Patterson~\cite{Pouq:num} arrived at a similar
conclusion based on their numerical simulations.  Their simulation
results also indicated that magnetic helicity, not necessarily the
residual helicity, is important for the growth of large-scale magnetic
field. Earlier, Batchelor \cite{Batc:MHD} had argued for transfer
between kinetic and magnetic energies at small scales; his arguments
were based on an analogy between the magnetic field equation of MHD
and the vorticity equation.

In a 2D EDQNM study, Pouquet \cite{Pouq:EDQNM2D} obtained eddy
viscosities for MHD turbulence. Upon injection of kinetic and magnetic
energy, she obtained a quasi stationary-state with a direct cascade of
energy to small scales together with an inverse cascade of mean-square
magnetic vector potential. In the inverse cascade regime, she argued
that the small-scale ME acts like a negative eddy viscosity on the
large-scale ME. The inverse cascade of the mean-square magnetic vector
potential was conjectured to arise due to the destablization of the
large-scale magnetic field by the small-scale magnetic field. She also
argued that the small-scale KE has the effect of a positive eddy
viscosity on the large-scale ME.  Ishizawa and Hattori
\cite{Ishi:EDQNM} in their EDQNM calculation obtained the eddy
viscosity due to each of the nonlinear terms in MHD equations.  They
found that the eddy viscosity due to ${\bf (b.\nabla)u}$ is positive,
leading to a transfer of energy from large-scale magnetic field to
small-scale magnetic field, apparently contradicting Pouquet's work
\cite{Pouq:EDQNM2D}.  However, this discrepancy may be because Pouquet
considers the regime of inverse cascade of mean-square vector
potential, while Ishizawa and Hattori \cite{Ishi:EDQNM} calculation
may capture the other wavenumber range.  One of the similarities
between Pouquet \cite{Pouq:EDQNM2D} and Ishizawa and Hattori's
\cite{Ishi:EDQNM} calculations is that both of them give a non-local
energy transfer from small-scale velocity field to the large-scale
velocity field.  In a recent work Ishizawa and Hattori
\cite{Ishi:flux} employed the wavelet basis to investigate energy
transfer in 2D MHD turbulence.

In this paper we have computed various cascade rates and other forms
of energy transfers using 2D MHD simulations.  It is in anticipation
that some of the conclusion drawn here will shed light on the nature
of interactions of both 2D and 3D MHD because the nature of cascade of
the total energy is similar in 2D and 3D \cite{Mont:SW}.  It is known
that the ME decays in 2D MHD turbulence \cite{Zeld:book}.  However,
this decay occurs only after a finite time, during which the ME is
steady.  We have calculated our fluxes and shell-to-shell energy
transfer rates for this ``quasi steady state''. We must however add
that the methods used here are completely generalizable to the 3D
case.  We have restricted ourselves to 2D purely because of
unavailability of powerful computing resources to us.

Most of the earlier work on energy transfer in MHD turbulence (e.g.,
\cite{Pouq:EDQNM,Pouq:num}) often dealt mainly with coarse-grained
energy transfer (between large scales and small scales). In a recent
work, Frick and Sokoloff \cite{Fric} solved a shell model of MHD
turbulence and calculated only the KE flux between the
velocity modes, and the ME flux between the magnetic
modes.  In our simulations we investigate energy transfer between {\bf
u}-to-{\bf u}, {\bf b}-to-{\bf b}, and {\bf u}-to-{\bf b} modes.
These results provide us with more informed picture of the physics of
energy transfer in MHD turbulence.

The paper is organized as follows: In Section \ref{s:formalism} we
discuss the known results of energy transfer in a MHD triad, and
introduce a new formalism called ``mode-to-mode energy transfer''.
Using the formalism of the mode-to-mode transfer, we derive in Section
\ref{s:flux_shell_to_shell} the formulas for various energy fluxes and
shell-to-shell energy transfer rates.  The simulation methodology is
discussed in Section \ref{s:sim_details}.  In Section \ref{s:results}
we report the numerically calculated energy fluxes and shell-to-shell
transfer rates.  The results are summarized in the last section.
Appendix contains derivation of the formulas for the mode-to-mode
transfer rates.

\section{Energy transfers in a MHD triad}
\label{s:formalism}

We will first state the known result on the energy transfers in a
triad.  Then we will introduce a new formalism called ``mode-to-mode''
energy transfer in turbulence.

The incompressible MHD equations in real space are 
\eqnMHDubx 
where ${\bf u}$ and ${\bf b}$ are the velocity and magnetic fields
respectively, $p$ is the total pressure, and $\nu$ and $\eta$ are the
kinematic viscosity and magnetic diffusivity respectively.  We have
assumed that the mean magnetic field is zero.  In Fourier space, the
KE ($E^{u}({\bf k})={\bf |u(k)|}^{2}/2$) and the ME
($E^{b}({\bf k})={\bf |b(k)|}^{2}/2$) evolution equations are
\cite{Pouq:EDQNM,Stan:book,Lesl:book}
\bea
\frac{\partial E^{u}(\bf k)}{\partial t} + 2 \nu k^{2} E^{u}({\bf k})&  = &
\sum_{{\bf k+p+q}=0} \frac{1}{2}S^{uu}({\bf k|p,q}) + 
	\sum_{{\bf k+p+q}=0} \frac{1}{2}S^{ub}({\bf k|p,q})  
	\label{eqn:eu_mhd}  \\
\frac{\partial E^{b}(\bf k)}{\partial t} + 2 \eta k^{2} E^{b}({\bf k}) & = &
\sum_{{\bf k+p+q}=0} \frac{1}{2}S^{bb}({\bf k|p,q}) +
	\sum_{{\bf k+p+q}=0} \frac{1}{2}S^{bu}({\bf k|p,q}) 
	\label{eqn:eb_mhd}
\eea
The four nonlinear terms  $S^{uu}({\bf k|p,q})$, $S^{ub}({\bf k|p,q})$,
$S^{bb}({\bf k|p,q})$, and $S^{bu}({\bf k|p,q})$ are 
\bea
      S^{uu}({\bf k|p,q}) & = & {{\slash{S}}}^{uu}({\bf k|p|q}) +
				{{\slash{S}}}^{uu}({\bf k|q|p}) 
		\label{eqn:Suu} \\
      S^{bb}({\bf k|p,q}) & = & {\cal{\slash{S}}}^{bb}({\bf k|p|q}) +
				{\cal{\slash{S}}}^{bb}({\bf k|q|p}) 
		\label{eqn:Sbb}  \\
      S^{ub}({\bf k|p,q}) & = & {\cal{\slash{S}}}^{ub}({\bf k|p|q}) +
				{\cal{\slash{S}}}^{ub}({\bf k|q|p}) 
		\label{eqn:Sub}  \\
      S^{bu}({\bf k|p,q}) & = & {\cal{\slash{S}}}^{bu}({\bf k|p|q}) +
				{\cal{\slash{S}}}^{bu}({\bf k|q|p})
		\label{eqn:Sbu}
\eea
where
\bea
{\cal{\slash{S}}}^{uu}({\bf k|p|q})  & = & 
		 -  \Im ({\bf [k.u(q)][u(k).u(p)] }) 
		\label{eqn:Suu_mode}  \\
{\cal{\slash{S}}}^{bb}({\bf k|p|q})  & = & 
		 -  \Im ({\bf [k.u(q)][b(k).b(p)] }) 
		\label{eqn:Sbb_mode}  \\
{\cal{\slash{S}}}^{ub}({\bf k|p|q})  & = & 
		  \Im ({\bf [k.b(q)][u(k).b(p)] }) 
		\label{eqn:Sub_mode}  \\
{\cal{\slash{S}}}^{bu}({\bf k|p|q})  & = & 
		  \Im ({\bf [k.b(q)][b(k).u(p)] }) 
		\label{eqn:Sbu_mode}
\eea
where $\Im$ stands for the imaginary part of the argument.

The terms $S^{\beta \alpha}({\bf k|p,q})$, where $\alpha,\beta$ stand
for $u$ or $b$, are conventionally taken to represent the nonlinear
energy transfer from modes ${\bf p}$ and ${\bf q}$ to mode ${\bf k}$
\cite{Pouq:EDQNM,Stan:book,Lesl:book}. In the following discussion we
derive a more detailed expression to obtain the energy transfer
between any two modes within a triad; we will refer to this as the
{\it ``mode-to-mode transfer''}. We emphasize that this approach is still
within the framework of the triad interaction, i.e., the triad is
still the fundamental entity of interaction of which the mode-to-mode
energy transfer is a part.  An expression for the energy transfer
between two modes of a triad by the mediation of the third mode is
sought here.

\begin{figure}[h]
\centerline{\mbox{\psfig{file=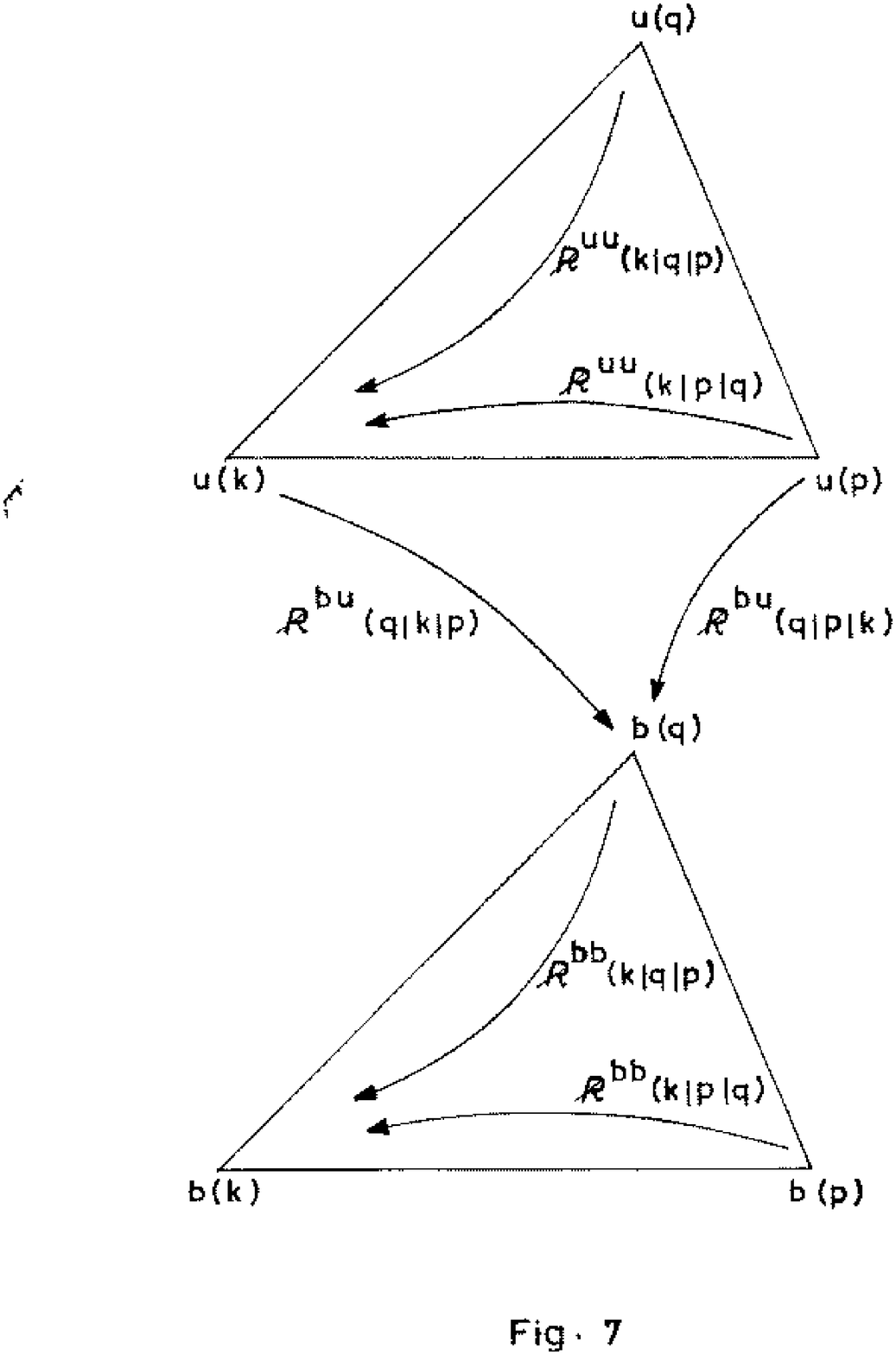,width=0.45\textwidth}}}
\caption{}
\label{fig:modetomode}
\end{figure}

Suppose in the triad (${\bf k,p,q}$), ${\cal{\slash{R}}}^{uu}({\bf
k|p|q})$ denotes the KE transfer from mode ${\bf p}$ to mode ${\bf k}$
with mode ${\bf q}$ as a mediator, and ${\cal{\slash{R}}}^{uu}({\bf
k|q|p})$ denotes the KE transfer from mode ${\bf q}$ to mode ${\bf k}$
with ${\bf p}$ as a mediator, then by definition,
\be
{\cal{\slash{R}}}^{uu}({\bf k|p|q}) + {\cal{\slash{R}}}^{uu}({\bf k|q|p})
                                    = S^{uu}({\bf k|p,q}) .
\ee
In addition, the energy transfer from {\bf u(k}) to {\bf u(p)},
${\cal{\slash{R}}}^{uu}({\bf k|p|q})$, 
should be equal and opposite to the transfer from {\bf
u(p}) to {\bf u(k)}, ${\cal{\slash{R}}}^{uu}({\bf p|k|q})$, that is, 
\be 
{\cal{\slash{R}}}^{bb}({\bf k|p|q})+ 
{\cal{\slash{R}}}^{bb}({\bf p|k|q})=0. 
\ee
Similar equations can be written for other pair of modes, and also for
$b$-to-$b$ and $u$-to-$b$ transfer (see Fig.~\ref{fig:modetomode}).
Our objective is to obtain an expression for ${\cal{\slash{R}}}^{\beta
\alpha}({\bf k|p|q})$ in terms of ${\bf k,p,q, u(k), u(p), u(q), b(k),
b(p), b(q)}$.

It is shown in the Appendix that ${\cal{\slash{R}}}^{uu}({\bf
k|p|q}) = {\cal{\slash{S}}}^{uu}({\bf k|p|q}) + X_{\Delta}$;
${\cal{\slash{R}}}^{bb}({\bf
k|p|q}) = {\cal{\slash{S}}}^{bb}({\bf k|p|q}) + Y_{\Delta}$,
and
${\cal{\slash{R}}}^{bu}({\bf
k|p|q}) = {\cal{\slash{S}}}^{bu}({\bf k|p|q}) + X_{\Delta}$,
where ${\cal{\slash{S}}}^{\beta \alpha}({\bf k|p|q})$s are given by
Eq.~(\ref{eqn:Suu_mode}-\ref{eqn:Sbu_mode}).  In Fig.~\ref{fig:S} we
illustrate all these mode-to-mode transfers ${\cal{\slash{R}}}^{\alpha
\beta}({\bf k|p|q})$.  The constants $X_{\Delta}, Y_{\Delta},
Z_{\Delta}$ have special significance, and are described below.

We take KE transfer rate as an example.  In Fig.~\ref{fig:S} we
observe that ${\cal{\slash{S}}}^{uu}({\bf k|p|q}) + X_{\Delta}$ gets
transferred from {\bf u(p)} to {\bf u(k)},
${\cal{\slash{S}}}^{uu}({\bf q|k|p}) + X_{\Delta}$ gets transferred
from {\bf u(k)} to {\bf u(q)}, and ${\cal{\slash{S}}}^{uu}({\bf p|q|k}) +
X_{\Delta}$ gets transferred from {\bf u(q)} to {\bf u(p)}.  The
quantity $X_\Delta$ flows along ${\bf u(p)} \rightarrow {\bf u(k)}
\rightarrow {\bf u(q)} \rightarrow {\bf u(p)}$, circulating around the
entire triad without changing the energy of any of the velocity
modes. Therefore we will call it the {\it ``circulating transfer''}.
The circulating transfer  lost by a mode is regained, hence, this
term does not affect the overall energy input/output of a mode.
Therefore ${\cal{\slash{S}}}^{uu}({\bf k|p|q})$
[cf. Eq.~(\ref{eqn:Suu_mode})] is the only term which effectively
participates in the energy transfer.  That is why it is termed as the
{\it ``effective mode-to-mode transfer rate''} from mode {\bf u(p)} to mode
{\bf u(k)}, mediated by mode {\bf u(q)}.  There are two other
circulating transfers: $Y_{\Delta}$ for $b$-to-$b$ transfer, and
$Z_{\Delta}$ for $u$-to-$b$ transfer; they are illustrated in
Fig.~\ref{fig:S}. A derivation of the effective mode-to-mode transfer
is given in Appendix.  For further details, refer to Ph.~D. thesis of
Dar \cite{Dar:thesis} and Dar et al.~\cite{Dar:modetomode}.

In summary, we have obtained a formula for ``effective mode-to-mode
energy transfer rates'' in MHD turbulence.  The effective KE transfer
rates from mode ${\bf p}$ to ${\bf k}$ is ${\cal{\slash{S}}}^{uu}({\bf
k|p|q})$, the effective ME transfer rates from mode ${\bf p}$ to ${\bf
k}$ is ${\cal{\slash{S}}}^{bb}({\bf k|p|q})$, and the effective
``conversion'' rate of KE of mode ${\bf p}$ to ME of mode ${\bf k}$ is
${\cal{\slash{S}}}^{bu}({\bf k|p|q})$.  The rate
${\cal{\slash{S}}}^{bu}({\bf k|p|q})$ is responsible for the growth of
magnetic energy.  In all these effective energy transfers, the mode
with wavenumber ${\bf q}$ {\it mediates} the transfer.

In the next section we will use our ``effective mode-to-mode energy
transfer rate'' formalism to derive formulas for 
cascade rates and shell-to-shell energy transfer.

\section{Cascade rates and Shell-to-Shell energy transfer in MHD turbulence}
\label{s:flux_shell_to_shell}

In this section we will use the formalism of effective mode-to-mode
transfers to define cascade rates and shell-to-shell energy transfer
rates in MHD turbulence.  Note that the formulas for cascade rate
etc. derived using our mode-to-mode transfer is equivalent to those
derive with $S(k|p,q)$.  However, our formalism has certain
advantages.  For example, some of the cascade rates and shell-to-shell
transfers (defined below) which were not accessible with the earlier
formalisms can now be calculated using our scheme.  In addition, the
flux and shell-to-shell transfer formulas using our schemes are
simpler.  In this paper we shall use the term {\it u}-sphere (shell)
to denote a sphere (shell) wavenumber-space containing
velocity modes.  Similarly, the {\it b}-sphere (shell) 
denote the corresponding sphere (shell) of magnetic modes.

\subsection{Energy cascade rates}
\label{subs:flux}

The energy cascade rate (or flux) is defined as the rate of loss of
energy from a sphere in the wavenumber space to the modes outside.
There are various types of cascade rates in MHD turbulence.  We have
schematically shown these transfers in Fig.~\ref{fig:flux}.  The
energy transfer could take take place from inside/outside $u/b$-sphere
to inside/outside-$u/b$ sphere.  In terms of
${\cal{\slash{S}}}^{\alpha \beta}({\bf k|p|q})$, the energy cascade
rate from inside of the $\alpha$-sphere of radius $K$ to outside of
the $\beta$-sphere of the same radius is
\be
\Pi^{\alpha<}_{\beta>}(K) = \sum_{|{\bf k}|>K} \sum_{|{\bf p}|<K} 
            {\cal{\slash{S}}}^{\beta \alpha}({\bf k|p|q}) .
\label{eqn:flux_inout}
\ee 
where $\alpha$ and $\beta$ stand for $u$ or $b$.

There is a transfer of energy from a {\it u}-sphere of radius $K$ to
the {\it b}-sphere of the same radius. We can obtain the effective
flux from the {\it u}-sphere to the {\it b}-sphere using
\be
\Pi^{u<}_{b<}(K) =
   \sum_{|{\bf p}|<K} \sum_{|{\bf k}|<K} 
            {\cal{\slash{S}}}^{bu}({\bf k|p|q}) .
\label{eqn:flux_uinbin}
\ee
Similarly, the energy flux from modes outside of the {\it u}-sphere to
the modes outside of the {\it b}-sphere can be obtained by
\be
\Pi^{u>}_{b>}(K) =
   \sum_{|{\bf p}|>K} \sum_{|{\bf k}|>K} 
            {\cal{\slash{S}}}^{bu}({\bf k|p|q}) .
\label{eqn:flux_uoutbout}
\ee
The total effective flux is defined as the total energy (kinetic+magnetic) lost
by the 
$K$-sphere to the modes outside, i.e.,
\begin{equation}
\Pi_{tot}(K) = \Pi^{u<}_{u>}(K) + \Pi^{b<}_{b>}(K) + \Pi^{u<}_{b>}(K) +
                \Pi^{b<}_{u>}(K). 
\label{eq:flux_tot}
\end{equation}

A schematic illustration of the effective fluxes defined in
Eqs.~(\ref{eqn:flux_inout})-(\ref{eqn:flux_uoutbout}) is shown in
Fig.~\ref{fig:flux}.  In Section \ref{subs:flux_results} we will
present the values of these fluxes calculated using numerical simulations
of 2D MHD turbulence.

\subsection{Shell-to-Shell energy transfer rates}
\label{subs:shell_to_shell}

In this section we derive an expression for the energy transfer rates
between two shells.  Consider KE exchange between shell {\it m} and
shell {\it n} shown in Fig.~\ref{fig:shelltoshell}.  For the triads
shown in the figure, the mode {\bf k} is in shell {\it m}, the mode
{\bf p} is in shell {\it n}, and mode {\bf q} could be inside or
outside the shells. In terms of the mode-to-mode transfer
${\cal{\slash{R}}}^{uu}({\bf k|p|q})$, the rate of energy transfer
from the $n^{th}$ $u$-shell to the $m^{th}$ $u$-shell is defined as
\begin{equation}
T^{uu}_{mn} = \sum_{{\bf k} {\large \epsilon}{\it m}} 
	      \sum_{{\bf p} {\large \epsilon {\it n}}} 
		{\cal{\slash{R}}}^{uu}({\bf k|p|q})
\label{eq:shell_new_defn}
\end{equation}
where the {\bf k}-sum is over {\it u}-shell {\it m} and the {\bf
p}-sum is over {\it u}-shell {\it n} with {\bf k+p+q=0}.  
As discussed in the last section, the quantity
${\cal{\slash{R}}}^{uu}$ can be written as a sum of an effective
transfer ${\cal{\slash{S}}}^{uu}({\bf k|p|q})$ and a circulating
transfer $X_\Delta$.  We know from the last section that the
circulating transfer does not contribute to the energy change of
modes. From Fig.~\ref{fig:shelltoshell} we can see that $X_\Delta$
flows from shell {\it m} to shell {\it n} and then flows back to {\it
m} indirectly through the mode {\bf q}. Therefore the {\it effective}
energy transfer rate from $n^{th}$ $u$-shell to $m^{th}$ $u$-shell 
is 
\be 
T^{uu}_{mn} = \sum_{{\bf k} {\large \epsilon}{\it m}} 
			\sum_{{\bf p} {\large \epsilon {\it n}}} 
	{\cal{\slash{S}}}^{uu}({\bf k|p|q}).
\label{eqn:shelltoshell_uu}
\ee
A general formula for shell-to-shell transfer from $m^{th}$
$\alpha$-shell to $n^{th}$ $\beta$-shell is
\be 
T^{\beta \alpha}_{mn} = \sum_{{\bf k} {\large \epsilon}{\it m}} 
			\sum_{{\bf p} {\large \epsilon {\it n}}} 
	{\cal{\slash{S}}}^{\beta \alpha}({\bf k|p|q}),
\label{eqn:shelltoshell}
\ee

Earlier Domaradzki and Rogallo \cite{Doma:POF2}
had given the following formula for shell-to-shell energy transfer in terms
of  $S^{uu}({\bf k|p,q})$:
\be
T^{uu}_{mn} = \frac{1}{2}\sum_{{\bf k} {\large \epsilon}{\it m}} 
	\sum_{{\bf p} {\large \epsilon {\it n}}} S^{uu}({\bf k|p,q}).
\label{eqn:shell_old_defn}
\ee 
This formula is incorrect due to the following reasons.  The formula
(\ref{eqn:shell_old_defn}) has contributions from two types of
wavenumber triads (see Fig.~\ref{fig:shelltoshell}).  In both the
triads, the wavenumber ${\bf k}$ is located in the shell $m$.
However, Type I triads have both wavenumbers ${\bf p,q}$ in the shell
$n$, but Type II triads have only wavenumber ${\bf p}$ in the shell
$n$.  Clearly, to compute the energy transfer from shell $n$ to shell
$m$, we must include all the triad interactions of type I, but only
the mode {\bf p} to mode {\bf k} energy transfer of type II triads.
Inclusion of mode {\bf q} to mode {\bf k} will lead to incorrect
estimation.  This is the drawback of Domaradzki and Rogallo's
formalism.  The above inconsistency has been hinted at by Batchelor
\cite{Batc:MHD} and Domaradzki and Rogallo \cite{Doma:POF2}
themselves.

In summary, Batchelor's and Domaradzki and Rogallo's formalism does
not yield correct shell-to-shell energy transfers; we have used an
alternate formula [Eq.~(\ref{eqn:shelltoshell})] to compute them.  We
have numerically computed them using the numerical data of 2D MHD
turbulence.  These results, discussed in Section
~\ref{subs:shell_results}, provide important insights into the energy
transfer process in MHD turbulence.

\section{Simulation Details}
\label{s:sim_details}
\subsection{Numerical Method}
\label{subs:num_method}

   In our simulations we use the Els\"{a}sser variable ${\bf
z}^{\pm}={\bf u} \pm {\bf b}$ instead of {\bf u} and {\bf b}.  The MHD
equations (1) and (2) written in terms of ${\bf z}^{+}$ and ${\bf
z}^{-}$ are
\bea
\frac{\partial{\bf z}^{\pm}}{\partial{t}}  +
\left ({\bf z}^{\mp} \cdot {\bf \nabla}\right){\bf z}^{\pm} 
   & = & -{\bf \nabla}p +
 \nu_{\pm} \nabla^{2}{\bf z}^{\pm} + 
\nu_{\mp} \nabla^{2}{\bf z}^{\mp} \nonumber \\
 &  & + \left(\nu_{\pm}/k_{eq}^{2}\right)\nabla^{4}{\bf z}^{\pm} 
      + \left(\nu_{\mp}/k_{eq}^{2}\right)\nabla^{4}{\bf z}^{\mp} +{\bf F^{\pm}}
\label{eqn:z_mhd}
\eea 
where $\nu_{\pm}=(\nu \pm \eta)/2$.  In addition to the viscosity and
the magnetic diffusivity of Eqs.~(\ref{eqn:udot_x}) and
(\ref{eqn:bdot_x}), Eq.~(\ref{eqn:z_mhd}) includes hyperviscosity
($\nu_{\pm}/k_{eq}^{2}$) to damp out the energy at very high
wavenumbers.  We choose $\nu = \eta = 5 \times 10^{-6}$ for runs on a
grid of size $512 \times 512 $. The parameter $k_{eq}$ is chosen to be
14 for all the runs.  The terms ${\bf F^{\pm}}$ are the forcing
functions. The corresponding forcing functions for the $\bf u$ and the
$\bf b$ fields in terms of ${\bf F^{\pm}}$ will be ${\bf F}^{u}=({\bf
F}^{+}+{\bf F}^{-})/2$ and ${\bf F}^{b}=({\bf F}^{+}-{\bf
F}^{-})/2$. In our simulations, we do not force the magnetic field
(i.e, ${\bf F}^{b}=0$). Consequently, we get ${\bf F}^{+}={\bf
F}^{-}={\bf F}^{u}={\bf F}$.  The absence of magnetic forcing is
motivated by the dynamo (magnetic field amplification) calculations
where only KE is forced.  Some of the physical examples where only KE
is forced are galactic dynamo and earth's magnetic field generation.
Even though our simulations are two-dimensional, we have followed the
similar line of approach as the dynamo simulations.

We use the pseudo-spectral method \cite{Canu:book} to solve the above
equations in a periodic box of size $2 \pi \times 2 \pi$. In order to
remove the aliasing errors arising in the pseudo-spectral method a
square truncation is performed wherein all the modes with $|k_{x}|
\geq N/3$ or $|k_{y}| \geq N/3$ $(N=512)$ are set equal to zero. The
equations are time advanced using the second order Adam-Bashforth
scheme for the convective term and the Crank-Nicholson scheme for the
viscous terms.  The time step $\Delta t$ used for these runs is $5
\times 10^{-4}$.  All the quantities are non-dimensionalised using the
initial total energy (one unit) and length scale of $2 \pi$.  Normalized
cross-helicity, defined as
\begin{equation}
\sigma_{c} = \frac{2 \int {\bf u.b} d^{3}x}{\int (u^{2}+b^{2}) d^{3}x)} , 
\end{equation}
has the value approximately equal to 0.1 for the results discussed in the
paper.  However, we have carried out simulations up-to $\sigma_c
\approx 0.9$ and the results obtained for the higher $\sigma_c$'s were
found to be qualitatively similar to those discussed in this paper.

At each time step we construct $\bf F$ as a divergenceless
uncorrelated random function (${\bf \nabla. F}=0$). The $x$-component 
of $\bf F$ is determined at every time step by
\be
F_{x} = \left(\frac{k_{y}}{k}\right) \sqrt{\Delta t} {\cal F} e^{i \phi} ,
\ee
where ${\cal F}^{2}$ is equal to the average energy input rate per
mode, and phase $\phi$ is a uniformly distributed random variable
between $0$ and $2 \pi$.  The $y$-component of the forcing function is
obtained by using the divergenceless condition, which yields
\be
F_{y} = - \left(\frac{k_x}{k_y}\right) F_x.
\ee

We apply isotropic forcing over a wavenumber annulus
$4<k<5$.  The value of ${\cal F}^{2}$ is
determined from the average rate of the total energy input, which in
our simulation is chosen to be equal to 0.1.

\subsection{Numerical computation of fluxes}
\label{subs:num_flux}

To compute the fluxes we employ a method similar to that used by
Domaradzki and Rogallo \cite{Doma:POF2}.  We outline this method
below using $\Pi^{b<}_{u>}(K)$ as an example. In
Eq.~(\ref{eqn:flux_inout}) we substitute the expression for
${\cal{\slash{S}}}^{ub}(k|p|q)$ from Eq.~(\ref{eqn:Sub_mode}) :
\be
\Pi^{b<}_{u>}(K) =  \sum_{|{\bf k}|>K} \sum_{|{\bf p}|<K} 
                      \Im \left( {\bf \left[k.b(q)\right]} 
                                   {\bf \left[u(k).b(p)\right]} \right) 
\label{eqn:piblug}
\ee
A straightforward summation over {\bf k} and {\bf p} involves
$O(N^{2})$ ($N$ = grid size) operations, and would thus involve a
prohibitive computational cost for large $N$ simulations.  Instead,
the pseudo-spectral method can be used to compute the above flux in
$O(NlogN)$ operations. The procedure is as follows:

We define two ``truncated'' variables ${\bf u}^{>}$ and ${\bf b}^{<}$ 
as follows
\begin{equation}
 {\bf u}^{>}({\bf k}) = \left \{ \begin{array}{ll}
                              0            & \mbox{if $|{\bf k}|<K$} \\  
                          {\bf u}({\bf k}) & \mbox{if $|{\bf k}|>K$} 
                          \end{array}
                         \right.
\label{eqn:utrunc_def}
\end{equation}
and
\begin{equation}
{\bf b}^{<}({\bf p}) = \left \{ \begin{array}{ll}
                          {\bf b}({\bf p}) & \mbox{if $|{\bf p}|<K$}  \\
                              0            & \mbox{if $|{\bf p}|>K$} 
                          \end{array}
                         \right. 
\label{eqn:btrunc_def}
\end{equation}
Eq.~(\ref{eqn:piblug})
 written in terms of ${\bf u}^{>}$ and ${\bf b}^{<}$ reads as
follows
\begin{equation}
\Pi^{b<}_{u>}(K) =  \sum_{\bf k} \sum_{\bf p} 
                       \Im \left( {\bf \left[k.b(k-p)\right]} 
                              {\bf \left[u^{>}(k).b^{<}(p)\right]} \right) .
\label{eqn:flux_binuout_trunc1}
\end{equation}
The above equation may be written as
\begin{equation}
\Pi^{b<}_{u>}(K) =   \Im \left [ \sum_{\bf k}  k_{j} u_{i}^{>}({\bf k})
                \sum_{\bf p} b_{j}({\bf k-p}) b_{i}^{<}({\bf p}) \right ]
\label{eqn:flux_binuout_trunc2}
\end{equation}
The ${\bf p}$ summation in the above equation can be recognized as a
convolution sum. The right hand side of
Eq.~(\ref{eqn:flux_binuout_trunc2}) can be efficiently evaluated by
the pseudo-spectral method using the truncated variables ${\bf u}^{>}$
and ${\bf b}^{<}$.  This procedure has to be repeated for every value
of $K$ for which the flux needs to be computed. The rest of the fluxes
and transfer rates defined in Section \ref{s:flux_shell_to_shell} are
similarly computed.

In the following section we describe the results of our simulations.

\section{Simulation Results}
\label{s:results}

\subsection{Generation of Steady State}
\label{subs:steady_state}

We compute the fluxes and the shell-to-shell transfers for steady
state. However, the computational time required to obtain a
statistically steady state on a grid of size $512^{2}$ is large. So to
obtain a steady state in simulations on this grid, we proceed in
stages. First, we run on a grid of size $64^{2}$ till a steady state
is achieved. This steady state field is then used as the initial
condition to achieve a steady-state for $128^{2}$, and similarly to
the grid of sizes $256^{2}$ and $512^{2}$.

Theoretically, the magnetic energy in two-dimensional MHD decays in
the long run even with steady KE forcing \cite{Zeld:book}.  However,
we find that the ME remains steady for sufficiently long time before
it starts to decay.  For $128^2$ simulation, the decay of ME starts
only after $t=25-30$ time units, and for $512^2$, the decay starts
much later. We term this region of constant ME as ``quasi
steady-state''.  In Fig.~\ref{fig:energy} we show both KE and ME for
this state.  Here the Alfv\'{e}n ratio (KE/ME) fluctuates between the
values $0.4$ and $0.56$, hence ME dominates KE.  The flux and the
shell-to-shell transfer rates reported in this paper is averaged over
this quasi steady-state.  The averaging is done once in every unit of
non-dimensional time over 15 time units. The averaged spectra of KE
and ME over this period is shown in Fig.~\ref{fig:spectrum}.
For the intermediate range of wavenumbers (inertial range), the spectra
show a powerlaw behaviour.  The spectral indices are in the range of
1.5-1.7, hence the determination of MHD spectral index is quite
difficult.  However, from the flux studies, Verma et al.~\cite{MKV:MHDsimu}
had concluded that Kolmogorov's $5/3$ exponent is 
preferred over Kraichnan's $3/2$ exponent. M\"{u}ller and Biskamp 
\cite{Bisk1:Kolm} and Biskamp and M\"{u}ller arrived at a similar
conclusion based on their high resolution simulation.

\subsection{Flux studies}
\label{subs:flux_results}

We compute all the energy fluxes defined in subsection \ref{subs:flux}
by averaging over the outputs of our numerical simulation.  In
Fig.~\ref{fig:fluxvsk} we show all the fluxes. The total flux
$\Pi_{tot}$ is positive indicating that there is a net loss of energy
from the $K$-sphere to modes outside. In the wavenumber region
$25<K<50$, the total flux is approximately constant. This wavenumber
region is the inertial range.

The net transfer from KE to ME is a sum of the fluxes
$\Pi^{u<}_{b<}(K)$, $\Pi^{u<}_{b>}(K)$, $\Pi^{u>}_{b<}$, and
$\Pi^{u>}_{b>}(K)$.  We observe from Fig.~\ref{fig:fluxvsk} that the
fluxes $\Pi^{u<}_{b>}(K_{max})$, $\Pi^{u>}_{b<}(K_{max})$ and
$\Pi^{u>}_{b>}(K_{max})$ are zero, understandably so because there is
no mode outside this sphere of maximum radius $K_{max}$.  The flux
$\Pi^{u<}_{b<}(K_{max})$ is found to be positive, indicating that
there is a net transfer from kinetic energy to magnetic energy.

We remind the reader that the {\it u}-modes within a wavenumber sphere
are called the {\it u}-sphere, and the {\it b}-modes within the
wavenumber sphere are called the {\it b}-sphere.  We find that the
flux $\Pi^{u<}_{b<}(K)$ is positive---hence, KE is lost by a {\it
u}-sphere to the corresponding {\it b}-sphere. The flux
$\Pi^{u<}_{b>}(K)$ is also positive.  It means that a {\it u}-sphere
loses energy to the modes outside the {\it b}-sphere.  We find that
the flux $\Pi^{b<}_{u>}(K)$ is negative, which implies that the {\it
b}-sphere gains energy from modes outside the {\it u}-sphere.  Thus,
all these fluxes result in a transfer of KE to ME. However,
$\Pi^{u>}_{b>}(K)$ is negative, implying that there is some feedback
of energy from modes outside the {\it b}-sphere to modes outside the
{\it u}-sphere.  The {\it net} transfer, however, is from KE to ME.
We shall later show that the flux $\Pi^{u>}_{b>}(K)$ plays a crucial
role in driving the kinetic-to-kinetic flux $\Pi^{u<}_{u>}(K)$.

The energy gained by the {\it b}-spheres is found to be approximately
constant for the range $20 \leq K \leq K_{max}$. This can be seen from
the plot of $\Pi^{b<}_{u<}(K)+ \Pi^{b<}_{u>}$. The constancy of the
flux implies that a {\it b}-sphere of radius $K$ and that of radius of
$K+\Delta K$ get the same amount of energy from the {\it u}-modes.
Thus there is no {\it net} energy transfer from the {\it u}-modes into
a {\it b}-shell (of thickness $\Delta K$) of the inertial range.  We
therefore conclude that the {\it net} energy transfer from {\it
u}-modes to the {\it b}-sphere occurs within the $K \leq 20$ sphere
(the large length scales).

We find that the flux $\Pi^{u<}_{u>}(K)$ is negative, that is the
inertial range {\it u}-sphere gain energy from the outside {\it
u}-modes---this is consistent with the numerical simulations of
Ishizawa and Hattori \cite{Ishi:flux}.  This behaviour of kinetic
energy is known as an ``inverse cascade'' in literature and is
reminiscent of the inverse cascade of KE in 2D fluid turbulence
\cite{Lesl:book} and of mean square vector potential in 2D MHD
turbulence \cite{Pouq:EDQNM2D}.  Note, however that the KE in 2D MHD
turbulence is not an inviscid invariant. Fig.~\ref{fig:fluxvsk} shows
that the inverse cascade of kinetic energy exists in the wavenumber
range $K \le 60$. Even the modes that are being forced ($4 \le k \le
5$) gain energy from higher wavenumbers {\it u}-modes.  The source of
this energy is the positive flux $\Pi^{b>}_{u>}(K)$ from higher {\it
b}-modes to the higher {\it u}-modes.

We also observe from Fig.~\ref{fig:fluxvsk} that there is a loss of
magnetic energy from the {\it b}-sphere to the {\it b}-modes outside
[see $\Pi^{b<}_{b>}(K)$], i.e., the ME cascade is forward.  This
result is consistent with the numerical results of Ishizawa and
Hattori \cite{Ishi:flux}.  The flux $\Pi^{b<}_{b>}(K)$ is found to be
constant in the inertial range.  Using similar reasoning to that given
above, we can again conclude that the net magnetic energy transfer to
a wavenumber shell in the inertial range is zero.

The above mentioned results are schematically illustrates in
Fig.~\ref{fig:fluxnum} for an inertial range wavenumber $K=20$.  The
energy input due to forcing and the small inverse cascade
[$\Pi^{u<}_{u>}(K)$] into the {\it u}-sphere from higher {\it u}-modes
provides the energy input into the {\it u}-sphere. This energy is
transferred into and outside the {\it b}-sphere by $\Pi^{u<}_{b<}(K)$
and $\Pi^{u<}_{b>}(K)$ respectively, the latter transfer being the
most significant of all transfers.  The energy transferred into the
{\it b}-sphere from the {\it u}-sphere [$\Pi^{u<}_{b<}(K)$] and a
small input from the modes outside the {\it u}-sphere
[$-\Pi^{b<}_{u>}(K)$] cascades down to the higher wavenumber {\it
b}-modes. In the high wavenumber {\it b}-modes, the cascaded energy
[$\Pi^{b<}_{b>}(K)$] and the energy transferred from the {\it
u}-sphere are partly dissipated and partly fed back to the high
wavenumber {\it u}-modes [$\Pi^{b>}_{u>}(K)$]. The feedback to the
kinetic energy is mostly dissipated, but a small inverse cascade takes
some energy back into the {\it u}-sphere.  This is the qualitative
picture of the energy transfer in 2D MHD turbulence.

The net transfer to each of the four corners of the
Fig.~\ref{fig:fluxnum} sum to zero within the statistical error (which
is computed from the standard deviation of the sampled data). This is
consistent with a quasi-steady-state picture.  The results presented
here for the quasi-steady-state in a forced turbulence remain
qualitatively valid even for a decaying case---the direction of the
various fluxes for the decaying case are identical to that for the
forced simulation; but as the energy decays, the magnitudes of all the
fluxes decrease.

The fluxes give us information about the overall energy transfer from
inside/outside $u/b$-sphere to inside/outside $u/b$-sphere.  To obtain
a more detailed account of the energy transfer, energy exchange
between the wavenumber shells have been studied; these results are
presented in the next subsection.

\subsection{Shell-to-Shell energy transfer-rate studies}
\label{subs:shell_results}

Significant details of energy transfers are revealed by calculating
the shell-to-shell energy transfer rates $T^{\beta \alpha}_{mn}$
defined by Eq.~(\ref{eqn:shelltoshell}).  We partition the k-space
into shells at wavenumbers $k_{n} (n=1,2,3,...)
=1,16,19.02,22.62,...,2^{(n+14)/4}$.  The first shell extends from
$k_{1}=1$ to $k_{2}=16$, the second shell extends from $k_{2}=16$ to
$k_{3}=19.02$, ..., the $m^{th}$ shell extends from $k_{m}$ to
$k_{(m+1)}$.  Thus, the effective shell-to-shell energy transfer rate
from the $n^{th} {\it \alpha}$-shell to the $m^{th} {\it \beta}$-shell
[Eq.~(\ref{eqn:shelltoshell})] can be written as
\be
T^{\beta \alpha}_{mn} = \sum_{k_{m}<k<k_{m+1}} \sum_{k_{n}<p<k_{n+1}} 
               \sum_{\bf q}^{\Delta} {\cal{\slash{S}}}^{\beta \alpha}
				({\bf k|p|q}).
\label{eq:shell_utou}
\ee

The energy transfers ($u-u$, $b-b$, and $u-b$) involving shells which
are close in wavenumber space are called {\it local} transfers, as is
the convention. Two shells are considered close if the wavenumber
ratio of the larger shell to the smaller shell is less than 2 (in this
study this would include the shells between $n+4$ to $n-4$ for the
$n$th shell). Transfers involving shells more distant than the above
range are called {\it non-local}.  Since two fields are involved in
MHD, there are energy transfers between shells of same index. In the 
following discussions we will show that the energy transfer between
$u$ and $b$ shells of same index plays a major role in MHD.

In Figs.~\ref{fig:Tbb}, \ref{fig:Tuu}, and \ref{fig:Tbu} we plot the
energy transfer rates $T^{bb}_{mn}, T^{uu}_{mn}$ and $T^{bu}_{mn}$
vs. $m$ respectively for various values of $n$.  It is evident from
the figures that the transfer rates between shells in the inertial
range are virtually independent of the individual values of the
indices $m$ and $n$, and only dependent on their differences. This
means that the transfer rates in the inertial range are {\it
self-similar}. The differences in $T^{\beta \alpha}_{mn}$ for various
$n$ are smaller than the standard deviation of the sampled data,
indicating that the perceived self-similarity is statistically
significant.

In $T^{bb}_{mn}$ versus $m$ plot of Fig.~\ref{fig:Tbb} we find that
the transfer rates $T^{bb}_{mn}$ are negative for $m < n$ and they are
positive for $m > n$.  Hence a {\it b}-shell gains energy from the
{\it b}-shells of smaller wavenumbers and loses energy to the {\it
b}-shells of larger wavenumbers.  Hence, ME cascades from the smaller
wavenumbers to the higher wavenumbers (forward).  Since $T^{bb}_{mn}$
is self-similar, the energy lost from a shell ($n-\Delta n$) to $n$ is
equal to the energy lost from $n$ to the shell ($n+\Delta n$). Thus,
the net magnetic energy transfer into any inertial range shell is
zero.  In addition, we find that the most significant energy transfer
takes place among $n-1, n$ and $n+1$ shells.  Hence, $b-b$ energy
transfer is local.

In $T^{uu}_{mn}$ versus $m$ of Fig.~\ref{fig:Tuu} we find that the
most dominant transfers are from the $n^{th}$ {\it u}-shell to $(m=n
\pm 1)^{th}$ shells. From the sign of these transfers we see that KE
is gained from $(n-1)^{th}$ shell and lost to $(n+1)^{th}$ shell.
This means that the {\it local} transfers from the adjacent shells
result in a {\it forward} cascade of KE towards the {\it large}
wavenumbers. The transfers from other shells are largely negligible
except for the transfer to the shell $m=1$ (shown boxed on the left of
the figure), which represents a loss of KE from high wavenumber modes
to the $m=1$ shell.  This nonlocal transfer (to the first shell)
produces an inverse cascade of KE to the small wavenumbers, observed
earlier in Section~\ref{subs:flux_results}.  Thus $u-u$ energy
transfer has a forward energy cascade involving local interactions,
and an inverse cascade involving nonlocal interactions with the first
$u$-shell.  The latter dominates the former in 2D turbulence.

We now discuss our simulation results for the energy transfer rates
between {\it u}-shells and {\it b}-shells.  In Fig.~\ref{fig:Tbu} we
find that $T^{bu}_{mn}$ is positive for all $m$, except for $m=n-1$
and $n$. This implies that the $n^{th}$ {\it u}-shell loses energy to all
but $(n-1)^{th}$ and $n^{th}$ {\it b}-shells.  However, the energy gained by
$n^{th}$ and $(n-1)^{th}$ $b$-shell is larger than the total loss of energy
by the {\it u}-shell through all other transfers. Consequently, there
is a net gain of energy by the {\it u}-shells in the inertial
range. In Section~\ref{subs:flux_results} it was shown that
outside $b$-sphere loses energy to outside $u$-sphere; this energy
flux is primarily due to the above shell-to-shell transfer.

From Fig.~\ref{fig:Tbu} we also observe that $T^{bu}_{mn}$ for $m < n$
is smaller than that for $m > n$.  Consequently, there is a net
transfer of energy from a {\it u}-shell to higher wavenumber {\it
b}-shells.  From Fig.~\ref{fig:Tbu} we also find that the energy
transfer rate $T^{bu}_{1n}$ from the $n^{th}$ {\it u}-shell to the
$1^{st}$ {\it b}-shell is positive, implying that the $1^{st}$ {\it
b}-shell gains energy from the {\it u}-shells---this is a nonlocal
transfer of energy from the large {\it u}-modes to the small {\it
b}-modes. The negative flux $\Pi^{b<}_{u>}$ observed in
Section~\ref{subs:flux_results} (see Fig.~\ref{fig:fluxnum}) is due to
this nonlocal transfer.  Hence, the results on flux and the transfer
rates are consistent with each other.

In Fig.~\ref{fig:Tbun1} we show the transfer rates $T^{bu}_{mn}$ from
the $n=1$ {\it u}-shell to all the {\it b}-shells.  Comparing the
magnitudes of $T^{bu}_{mn}$ we notice that the energy transfer rate
from the $1^{st}$ {\it u}-shell dominates the transfers from all other
{\it u}-shells. Hence, there is a large amount of {\it non-local}
transfer from the first {\it u}-shell to the larger {\it b}-shells,
reminiscent of positive and large $\Pi^{u<}_{b>}$.  In
Fig.~\ref{fig:Tbusum} we plot the net KE transferred into a {\it
b}-shell (=$ \sum_{n} T^{bu}_{mn}$) ($\diamond$ in the figure), and
also KE transferred from all {\it u}-shells {\it except} the first
($+$ in the figure).  The figure clearly shows the net energy
transferred into the inertial range {\it b}-shell to be nearly zero,
but the KE transferred from the shells $n \ge 2$ is negative.  This
implies that the inertial range $b$-shells lose energy to the
$u$-modes of shells $n \ge 2$, but they gain roughly the equal amount
of energy from $n=1$ $u$-shell through nonlocal interaction.  The
overall transfer to the inertial range $b$-shells is negligible.  This
result is consistent with the picture obtained from the flux arguments
(Subsection \ref{subs:flux_results}, Para 4).

We schematically illustrate in Fig.~\ref{fig:T} the energy transfers
between shells.  In this figure we show directions of the most
significant energy transfers. The arrows indicate the directions of
the transfers, and thickness of the arrows indicates the approximate
relative magnitudes. Since the local transfer rates are self-similar,
the transfers from any other shells in the inertial range will also
show the same pattern. An inertial range {\it b}-shell gains energy
from the smaller {\it b}-shells ($T^{bb}_{mn}$) and the smaller {\it
u}-shells ($T^{bu}_{m1}$) through local and non-local transfers
respectively.  The energy gained by a {\it b}-shell from the smaller
{\it b}-shells is exclusively lost to the larger {\it b}-shells, and
the energy gained from the {\it u}-shells is mainly lost to the same
index $u$-shell by $T^{bu}_{nn}$.  Regarding the velocity modes, an
inertial range {\it u}-shell gains energy from smaller {\it u}-shells
by {\it local} transfers, and from the $b$-shell of same shell index.
An inertial range $u$-shell loses energy to higher $b$-shells (locally)
and to the first $u$-shell and the first $b$-shell (nonlocally).
  As illustrated in Fig.~\ref{fig:T}, there is a transfer
of energy from the first {\it u}-shell to the first {\it
b}-shell. This is one of the most significant gain of magnetic energy
from the kinetic energy. The first $u$-shell also loses energy to the
higher $b$-shell by nonlocal interactions.

To summarize, in the inertial range the forward magnetic energy
transfer and the forward kinetic energy transfers are local in nature.
There is also energy transfer from $b$-modes of shell index $n$ to the
$u$-modes of same indexed shell.  There is nonlocal transfer from
inertial range {\it u}-shells to the first {\it b}-shell, and from the
first {\it u}-shell to the {\it b}-shells. There is also a significant
energy transfer from the first {\it u}-shell to the first {\it
b}-shell.

In this section we described various cascade rates (fluxes) and
shell-to-shell energy transfer rates. It is clear that the complete
picture is quite complex.

\section{Conclusion}
\label{s:conclusion}

In literature we find a description of energy transfer rates from two
modes in a triad to the third mode, i.e., from {\bf u(p)} and {\bf
u(q)} to {\bf u(k)}. Here we have constructed new formulae to describe
energy transfer rates between a pair of modes in the triad
(mode-to-mode transfer), say from {\bf u(p)} to {\bf u(k)}, where the
third mode in the triad acts as a mediator in the transfer process.

In our formalism the mode-to-mode energy transfer can be expressed as
a combination of an ``effective transfer'' {\it and} a ``circulating
transfer''.  The circulating transfer will not result in a change of
modal energy, since the amount of circulating transfer gained by mode
{\bf k} from mode {\bf p} is also lost by mode {\bf k} to mode {\bf q}
(see Fig.~\ref{fig:S}).  Only the effective transfer is
responsible for modal energy change. As the circulating transfer does
not have any observable effect on the energy of the modes, it may be
correct to ignore it from the study of energy transfer.  Using the
notion of effective mode-to-mode transfer and the circulating
transfer, we defined effective cascade rates and shell-to-shell energy
transfer rates in MHD turbulence (see formulas of Section
\ref{s:flux_shell_to_shell}).  Some of these energy transfers can not be
calculated using the earlier formalisms, but they are accessible through
our  ``mode-to-mode'' energy transfer scheme.  

Some of our formulas are directly applicable to fluid turbulence.  For
example, the formulas for mode-to-mode energy transfer
${\cal{\slash{S}}}^{uu}({\bf k|p|q})$, the flux $\Pi^{u<}_{u>}$, and
the shell-to-shell energy transfer $T^{uu}_{mn}$ can be used for fluid
turbulence calculations without any alteration.

We investigated the features of kinetic and magnetic energy transfer
between various scales at low values of cross helicity in a
quasi-steady state of forced 2D MHD turbulence. Several interesting
observations were made in our simulations.  A summary of the results
is given below. For the following discussion refer to
Figs.~\ref{fig:fluxnum} and \ref{fig:T}.

\begin{enumerate}

\item  There is a net transfer of energy from the kinetic to the magnetic. 

\item There is an energy transfer to the large-scale magnetic field
from the large-scale velocity field ($\Pi^{u<}_{b<}$), and also from
the small-scale velocity field ($\Pi^{b<}_{u>}$).  The former transfer
is of a greater magnitude than the latter, hence the magnetic field
enhancement is primarily caused by a transfer from the {\it u}-sphere
to the {\it b}-sphere.  Indeed the first few {\it b}-modes get most of
this energy.  Other significant transfers between the velocity and
magnetic field are from the large-scale velocity field to the
small-scale magnetic field ($\Pi^{u>}_{b<}$) and an interesting
reverse transfer from the small-scale magnetic field to the
small-scale velocity field ($\Pi^{b>}_{u>}$).

\item The picture obtained from the shell-to-shell $u$-to-$b$ transfer is
more complex, but is consistent with the flux picture discussed in
item 2.  The first $u$-shell loses a large amount of energy to the
inertial range $b$-shells. On the contrary, the first $b$-shell gains
energy from the inertial range $u$-shells.  Both these processes are
nonlocal.  Also, the positivity of $\Pi^{u<}_{b>}$ and negativity of
$\Pi^{b<}_{u>}$ is mainly due to the above two shell-to-shell energy
transfers.  In addition to these interactions, an inertial range $n$th
$u$-shell  loses energy to all but $(n-1)^{th}$ and $n^{th}$ shells.
The quantity $T^{bu}_{mn}$  for $m<n$ is smaller than that for $m>n$,
hence the inertial range $u$-shells mainly lose energy to the higher
wavenumber $b$-shells.  The negative $\Pi^{u>}_{b>}$ is primarily due
to the energy transferred from the $n$th $b$-shell to $n$th $u$-shell.

\item There is a {\it forward} cascade of ME towards the
small scales. This is consistent with other recent numerical
simulations \cite{Ishi:flux}. EDQNM closure calculations also yield
a ME transfer to the small-scales \cite{Ishi:EDQNM}.

\item The shell-to-shell transfer between $b$-to-$b$ modes shows that
the most significant interaction of $n$-th $b$ shell takes place with
$(n \pm 1)$th $b$ shells; the $(n-1)$th $b$-shell loses energy to the
$n$th $b$-shell, which in turn loses energy to the $(n+1)$th shell.
Hence, the $b$-to-$b$ energy transfer is forward and local.

\item There is an {\it inverse} cascade in the velocity field---this
is consistent with the observation of Ishizawa and Hattori arising
from numerical simulations \cite{Ishi:flux} and is also consistent
with the EDQNM closure calculations \cite{Pouq:EDQNM2D,Ishi:EDQNM}.
This inverse cascade of KE is driven by the reverse
transfer of energy from magnetic to the velocity field at the small
scales. Although the flux study points to an inverse cascade of
kinetic energy, the shell-to-shell energy transfer rates reveal the
following feature. There exists both an inverse and a forward transfer
of kinetic energy. The inverse transfer is primarily {\it nonlocal},
coming from the large wavenumber $u$-modes to the small wavenumber
$u$-modes. The forward cascade is {\it local}, i.e., between same sized
eddies in the inertial range.

\item The shell-to-shell energy transfers between the inertial range
shells statistically depend only on the difference of the shell indices.
Hence, the inertial range interactions can be said to be self similar.

\end{enumerate}

As pointed out in the first section, Pouquet \cite{Pouq:EDQNM2D}
observed (based on her 2D EDQNM study) a forward cascade of energy and
an inverse cascade of mean-square vector potential, and Ishizawa and
Hattori \cite{Ishi:EDQNM} proposed that the large-scale magnetic
energy is enhanced due to energy transfer from the small-scale
velocity field to the large-scale magnetic field. We also find these
transfers in our 2D simulation, but the dominant transfer to the
large-scale ME is from the large-scale KE.  Our simulation can not
probe the inverse cascade region referred to by Pouquet
\cite{Pouq:EDQNM2D} because we force the small wavenumber modes, so
there is no scope of inverse cascade in our simulation.

Pouquet and Patterson \cite{Pouq:num} proposed in context of 3D
turbulence that there an inverse cascade of ME from small
length-scales to large length-scales in presence of helicity.  Since
there is no helicity present in 2D MHD, helical MHD is inaccessible to
our simulation.  Still we observe some very striking features in our
simulation.  For example, our numerical result regarding forward
cascade of ME ($\Pi^{b<}_{b>} > 0$) is in agreement with the forward
cascade of ME obtained by the perturbative calculation of nonhelical
MHD \cite{MKV:MHD_PRE,MKV:MHDflux}.  Similar analytic calculation in
presence of helicity is in progress.  Full three-dimensional
simulation and analytic work in this direction will be useful in
resolving various issues in the magnetic field generation problem.

The fluxes and the transfer rates discussed here and in the
analytic paper by Verma \cite{MKV:MHD_PRE,MKV:MHDflux} could
find applications in the dynamo problem. In astrophysical objects,
like galaxies and the sun, the magnetic field is thought to have
arisen due to the amplification of a seed magnetic field. Our study
has been performed over a quasi-steady state with a low Alfv\'{e}n
ratio. In order to understand the build-up of magnetic energy starting
from a seed value it is important to perform a similar study at high
Alfv\'{e}n ratio.  In some of the popular models, like the
$\alpha$-dynamo \cite{Zeld:book}, the mean magnetic field gets
amplified in presence of helical fluctuations.  It must be borne in
mind that our calculations are two-dimensional and devoid of magnetic
helicity and kinetic helicity. A three-dimensional calculation (with
the inclusion of magnetic and kinetic helicities) of various fluxes
and shell-to-shell energy transfer rates will yield important insights
into some unresolved issues concerning enhancement of magnetic energy.

The fluxes also find important applications in various phenomenological
studies. For example, Verma {\it et al.} \cite{MKV:SWheat} estimated the
turbulent dissipation rates in the solar wind and obtained the
temperature variation of the solar wind as a function of solar
distance. The various cascade rates discussed here could be useful
for various astrophysical studies. For example, $\Pi^{u<}_{b<}$ and
$\Pi^{u<}_{b>}$ can be used for studying the variation of $r_{A}$
of the solar wind.

The physics of MHD turbulence is still unclear.  The studies of
various fluxes and transfer rates shed light at various aspects which
will help us in getting a better understanding of MHD turbulence.

\acknowledgments

We thank Prof.~R.~K.~Ghosh of Computer Science Dept., Indian Institute
of Technology (IIT) Kanpur, for providing us computer time through the
project TAPTEC/COMPUTER/504 sponsored by All India Council for
Technical Education (AICTE).

\appendix
\section{Derivation of Mode-to-Mode energy transfers in MHD turbulence}

In this appendix we will derive the result stated in Section
\ref{s:formalism} that the ``effective '' energy transfer rate from
mode ${\bf p}$ to mode ${\bf k}$ (${\cal{\slash{R}}}^{\alpha
\beta}({\bf k|p|q})$) is equal to ${\cal{\slash{S}}}^{\alpha
\beta}({\bf k|p|q})$.  We will find in detail the mode-to-mode energy
transfer from {\bf u(p)} to {\bf u(k)} within the triad ({\bf
k,p,q}). The other mode-to-mode energy transfer from {\bf b(p)} to
{\bf b(k)} and from {\bf u(p)} to {\bf b(k)} can be obtained by
following the same procedure.  In all these mode-to-mode transfers,
mode ${\bf q}$ acts as a mediator.

\subsection{Definition of Mode-to-Mode transfer in a triad}
\label{subs:mode_to_mode_defn}

In the beginning we limit ourselves to $u$ to $u$ transfer in a triad
({\bf k, p, q}).  Let the quantity ${\cal{\slash{R}}}^{uu}({\bf k|p|q})$
denote the energy transferred from mode {\bf p} to mode {\bf k} with
mode {\bf q} playing the role of a mediator (see
Fig.~\ref{fig:modetomode}).  We wish to obtain an expression for
${\cal{\slash{R}}}^{uu}$.  

The ${\cal{\slash{R}}}^{uu}$ should satisfy the following relationships :
\begin{enumerate}
\item The sum of ${\cal{\slash{R}}}^{uu}({\bf k|p|q})$ and
${\cal{\slash{R}}}^{uu}({\bf k|q|p})$, which represent energy transfer
from mode {\bf p} to mode {\bf k} and from mode {\bf q} to mode {\bf
k} respectively, should be equal to the total energy transferred to
mode {\bf k} from modes {\bf p} and {\bf q}, i.e., $S^{uu}({\bf
k|p,q})$ [see Eq.~(\ref{eqn:eu_mhd})]. Thus,
\bea
{\cal{\slash{R}}}^{uu}({\bf k|p|q}) + {\cal{\slash{R}}}^{uu}({\bf k|q|p})
                                 & = & S^{uu}({\bf k|p,q}), 
		\label{eqn:Rkpq} 
\eea
Similarly,
\bea
{\cal{\slash{R}}}^{uu}({\bf p|k|q}) + {\cal{\slash{R}}}^{uu}({\bf p|q|k})
                                 & = & S^{uu}({\bf p|k,q}),  
		\label{eqn:Rpkq} \\
{\cal{\slash{R}}}^{uu}({\bf q|k|p}) + {\cal{\slash{R}}}^{uu}({\bf q|p|k})
                                 & = & S^{uu}({\bf q|k,p}).
		\label{eqn:Rqkp}
\eea

\item The energy transferred from mode {\bf p} to mode {\bf k}, 
i.e., ${\cal{\slash{R}}}^{uu}({\bf k|p|q})$, will be equal and opposite
to the energy transferred from mode {\bf k} to mode {\bf p}
i.e., ${\cal{\slash{R}}}^{uu}({\bf p|k|q})$. Thus,
\bea
{\cal{\slash{R}}}^{uu}({\bf k|p|q})+ {\cal{\slash{R}}}^{uu}({\bf p|k|q})
				& = & 0, 
			\label{eqn:R1}  \\
{\cal{\slash{R}}}^{uu}({\bf k|q|p})+ {\cal{\slash{R}}}^{uu}({\bf q|k|p})
				& = & 0 , 
			\label{eqn:R2}  \\
{\cal{\slash{R}}}^{uu}({\bf p|q|k})+ {\cal{\slash{R}}}^{uu}({\bf q|p|k})
				& = & 0 .
			\label{eqn:R3}
\eea
\end{enumerate}
These are six equations with six unknowns. However, the value of the
determinant formed from the
Eqs.~(\ref{eqn:Rkpq})-(\ref{eqn:R3}) is zero.  Therefore
we cannot find unique ${\cal{\slash{R}}}^{uu}$s given just these
equations.

\subsection{Solutions of equations of mode-to-mode transfer}
\label{subs:mode_to_mode_soln}

Consider 
\be
{\cal{\slash{S}}}^{uu}({\bf k|p|q}) \equiv 
	- \Im  \left( {\bf [k.u(q)] [u(k).u(p)]} \right) 
\label{eq:Sukup_def_appendix}
\ee

By definitions given in Eqs.~(\ref{eqn:Suu}),
${\cal{\slash{S}}}^{uu}$s satisfies
Eqs.~(\ref{eqn:Rkpq}-\ref{eqn:Rqkp}).  Using the triad relationship
${\bf k+p+q = 0}$, and the incompressibility constraint [${\bf
k.u(k)=0}$], it can be seen that ${\cal{\slash{S}}}^{uu}$s, etc. also
satisfy Eqs.~(\ref{eqn:R1}-\ref{eqn:R3}).  Hence, the set of
${\cal{\slash{S}}}^{uu}$s are {\it one instance} of the
${\cal{\slash{R}}}^{uu}$s, i.e., ${\cal{\slash{R}}}^{uu}({\bf
k|p|q})={\cal{\slash{S}}}^{uu}({\bf k|p|q})$.  However, this is not a
unique solution because the determinant formed from the
Eqs.~(\ref{eqn:Rkpq})-(\ref{eqn:R3}) is zero.

\begin{figure}[h]
\centerline{\mbox{\psfig{file=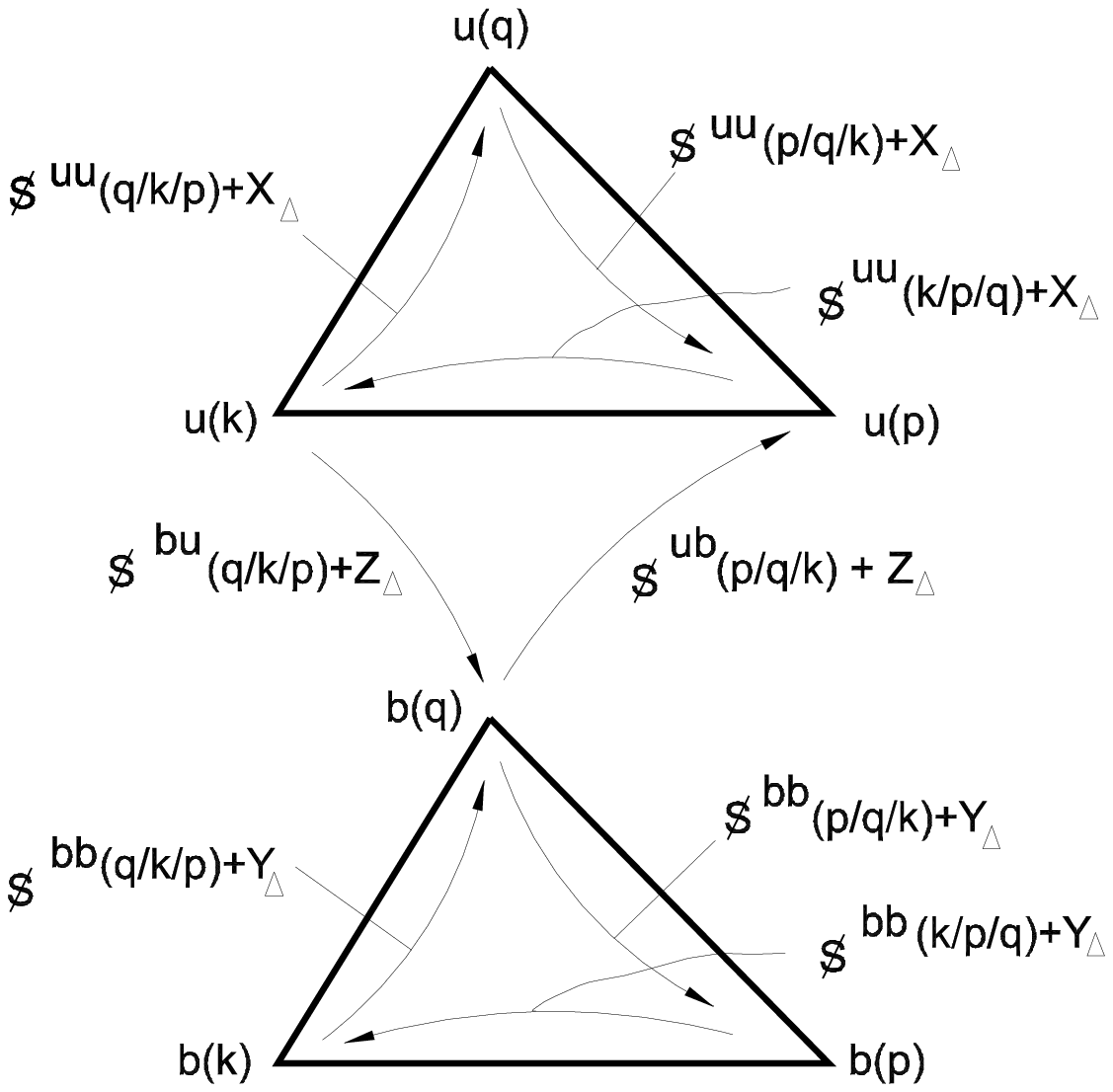,width=1.0\textwidth}}}
\caption{}
\label{fig:S}
\end{figure}

\begin{figure}[h]
\vskip -20cm
\psfig{file=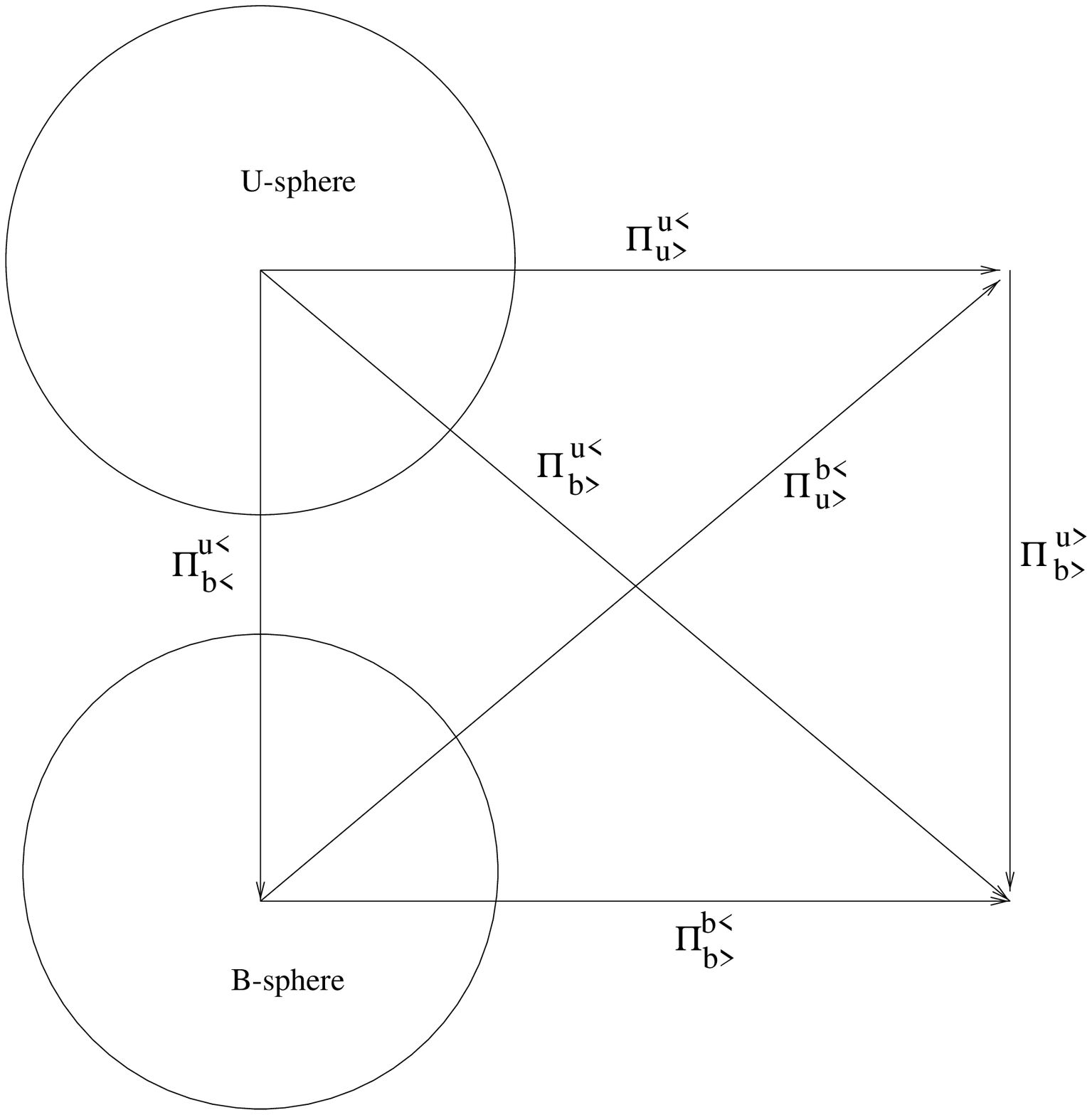,width=0.8\textwidth}
\caption{}
\label{fig:flux}
\end{figure}

\begin{figure}[h]
\centerline{\mbox{\psfig{file=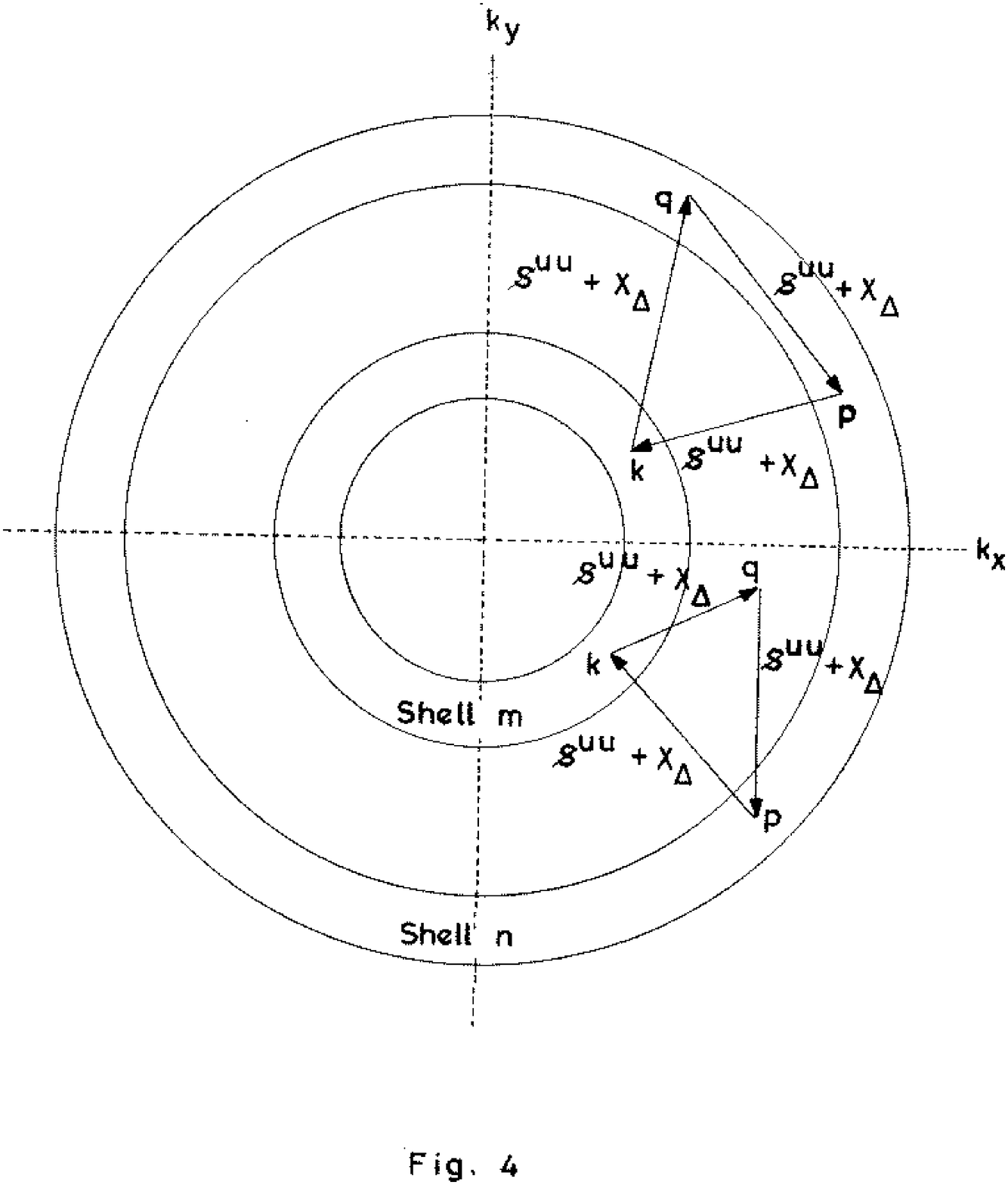,width=0.7\textwidth}}}
\caption{}
\label{fig:shelltoshell}
\end{figure}

\begin{figure}[h]
\vskip -20cm
\psfig{file=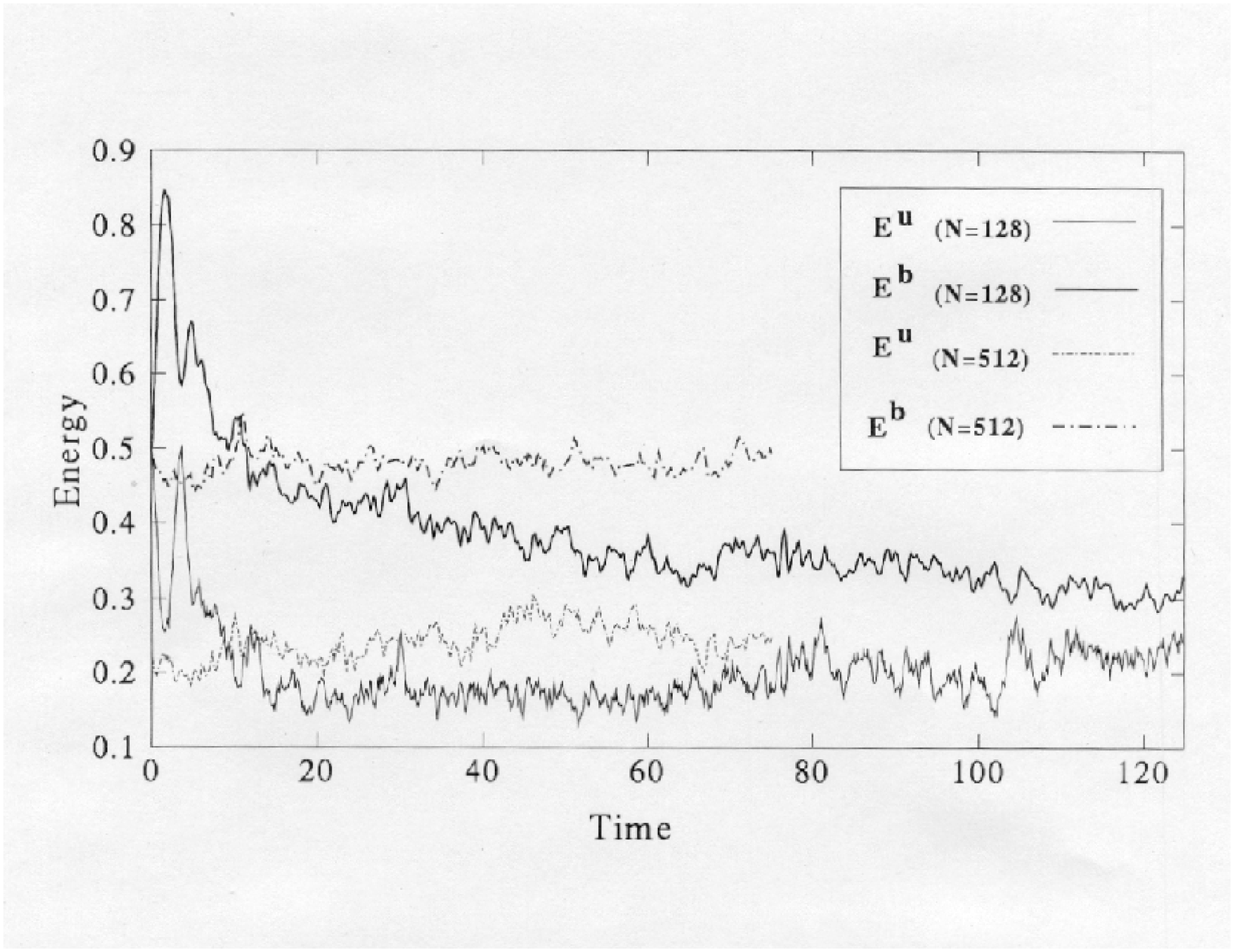,width=0.8\textwidth}
\caption{}
\label{fig:energy}
\end{figure}

\begin{figure}[h]
\vskip -20cm
\psfig{file=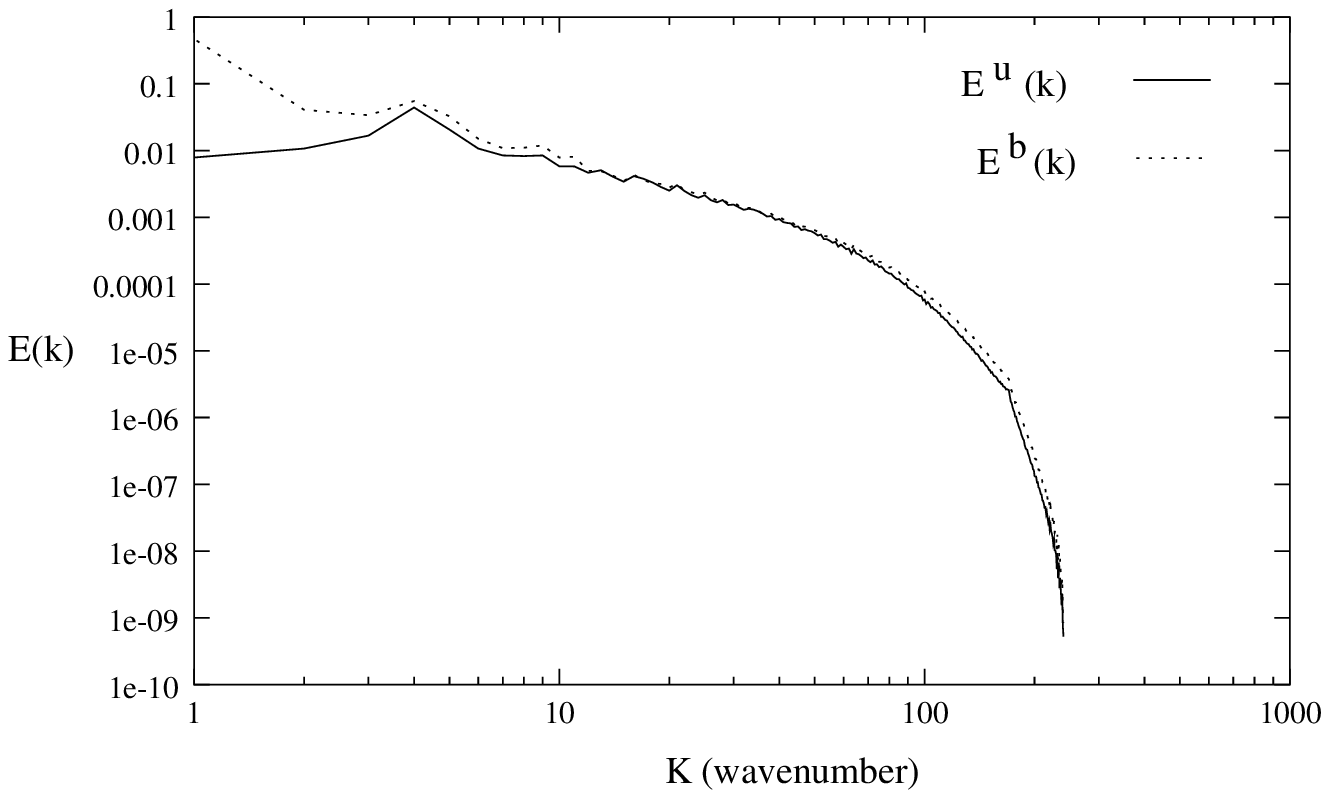,width=0.8\textwidth}
\caption{}
\label{fig:spectrum}
\end{figure}

\begin{figure}[h]
\vskip -20cm
\psfig{file=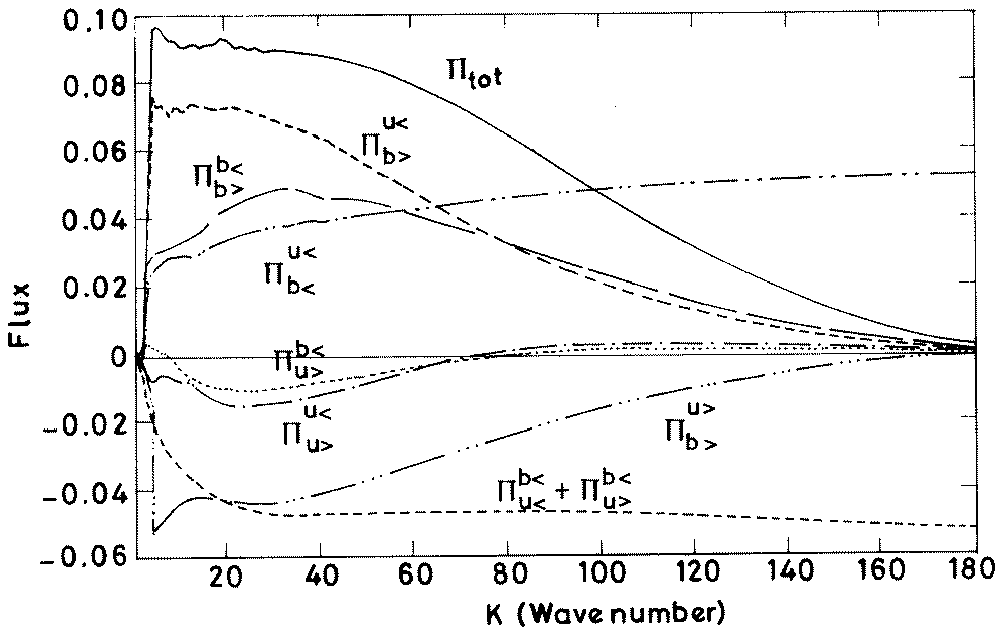,width=0.8\textwidth}
\caption{}
\label{fig:fluxvsk}
\end{figure}

\begin{figure}[h]
\centerline{\mbox{\psfig{file=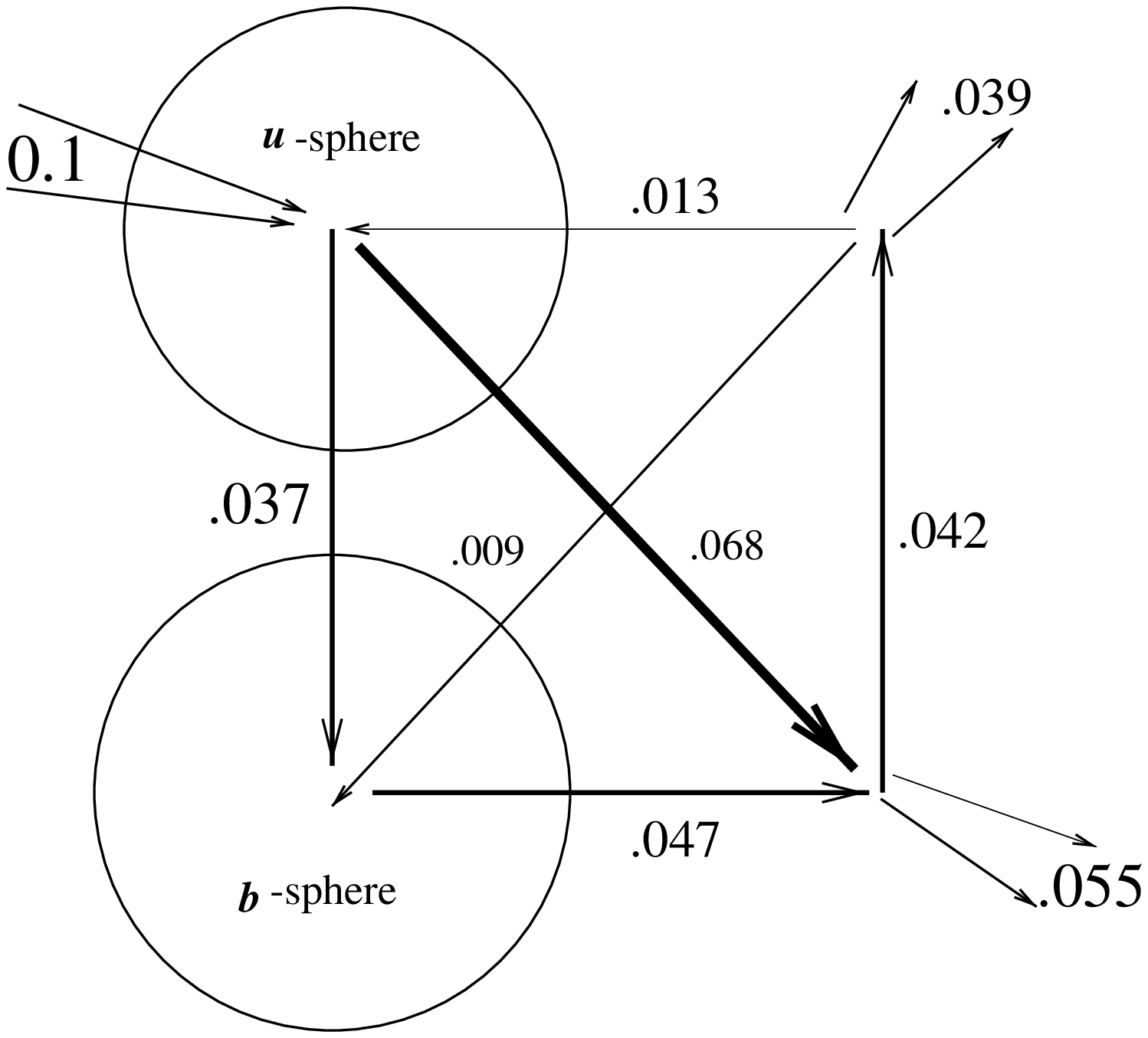,width=0.8\textwidth}}}
\caption{}
\label{fig:fluxnum}
\end{figure}

\begin{figure}[h]
\centerline{\mbox{\psfig{file=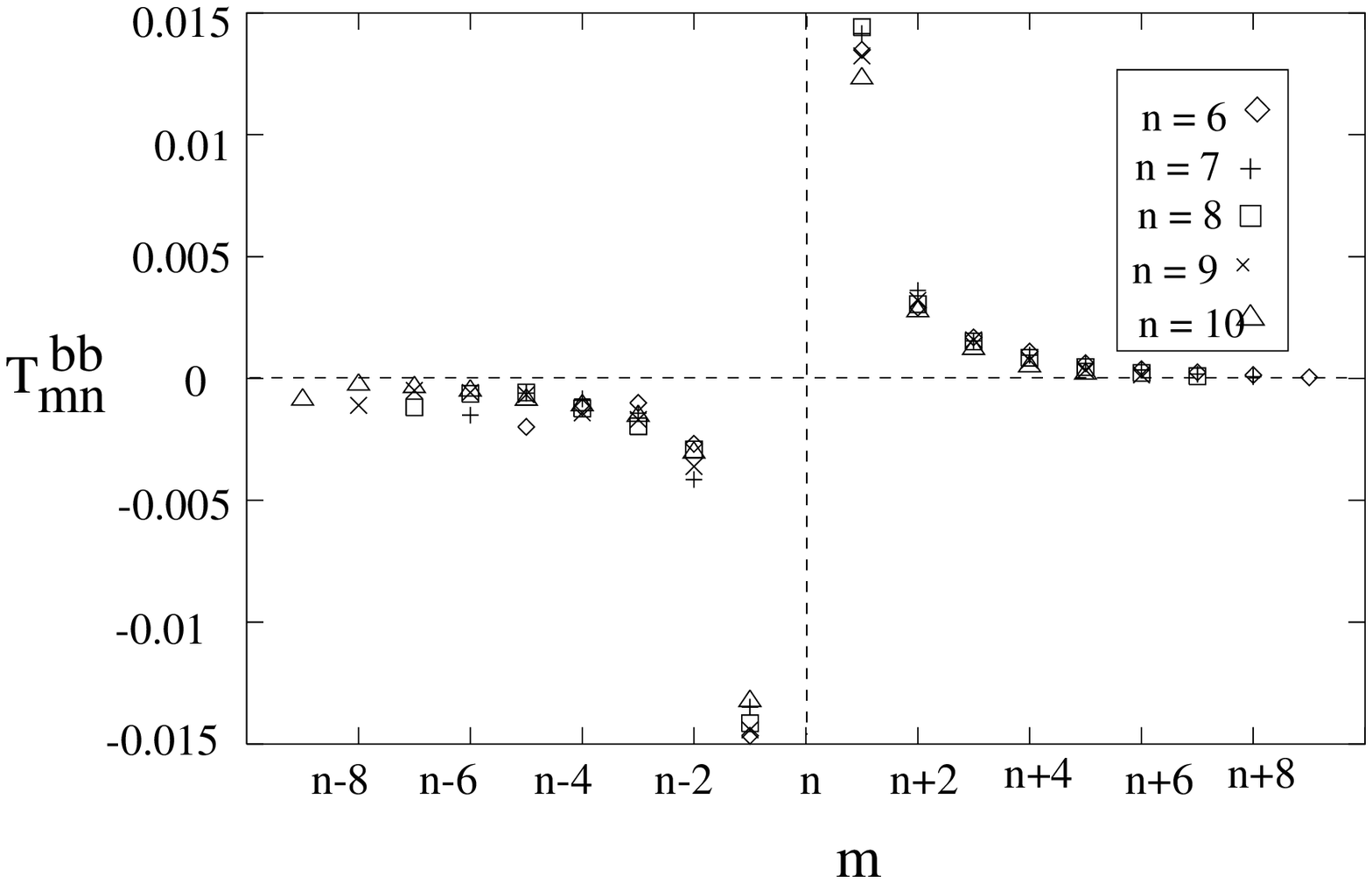,width=0.8\textwidth}}}
\caption{}
\label{fig:Tbb}
\end{figure}

\begin{figure}[h]
\centerline{\mbox{\psfig{file=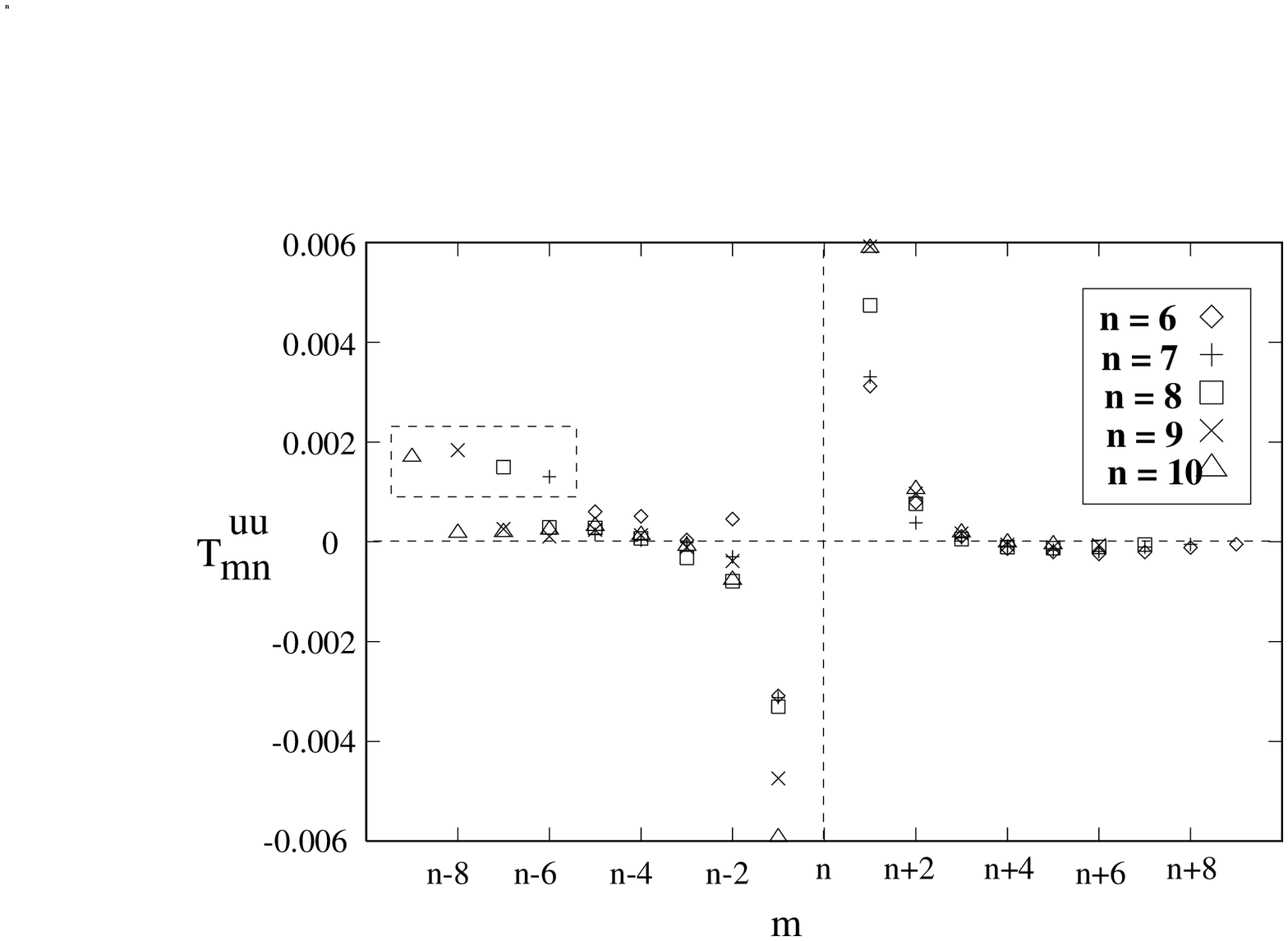,width=0.8\textwidth}}}
\caption{}
\label{fig:Tuu}
\end{figure}

\begin{figure}[h]
\centerline{\mbox{\psfig{file=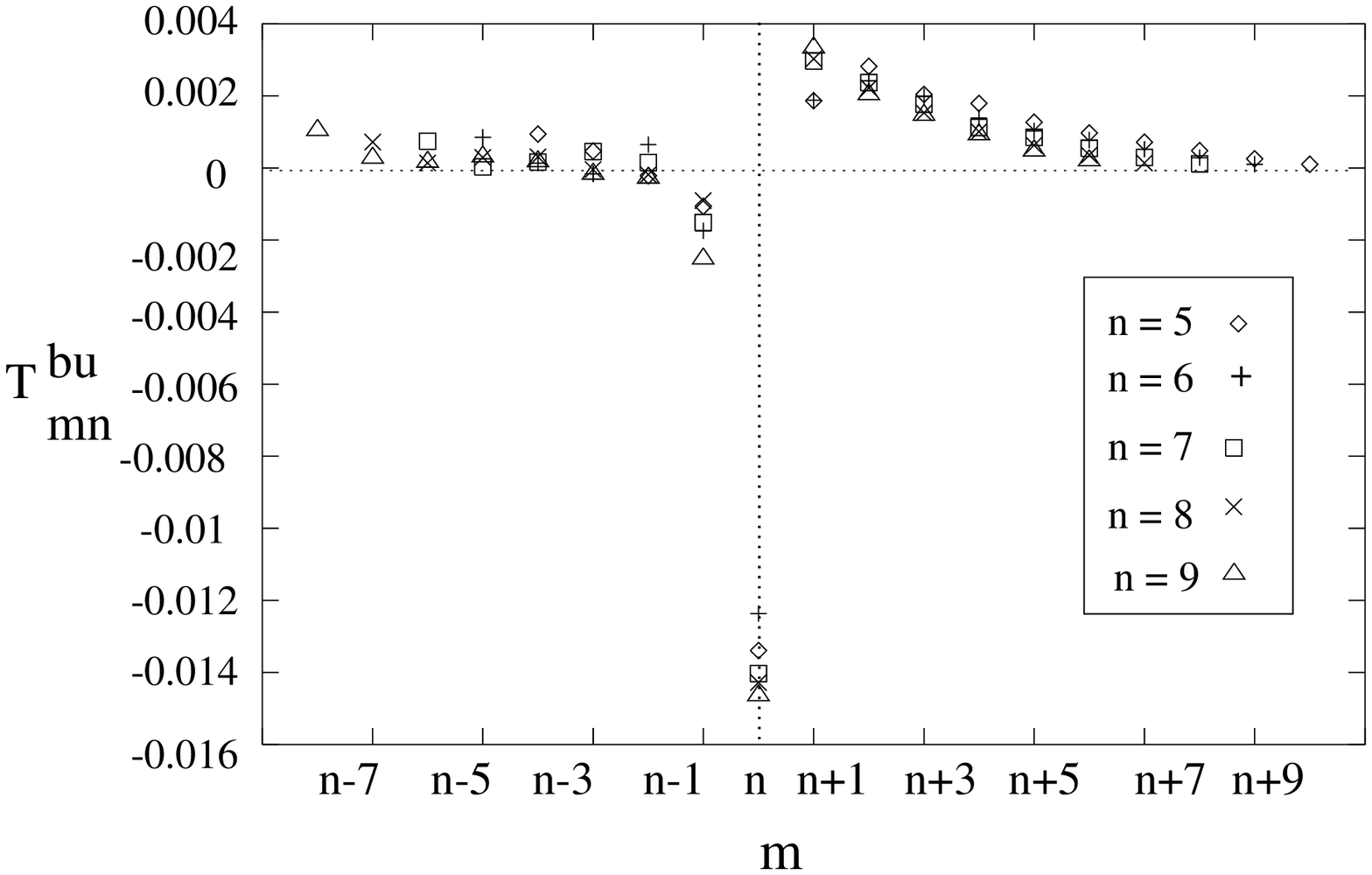,width=0.8\textwidth}}}
\caption{}
\label{fig:Tbu}
\end{figure}

\begin{figure}[h]
\centerline{\mbox{\psfig{file=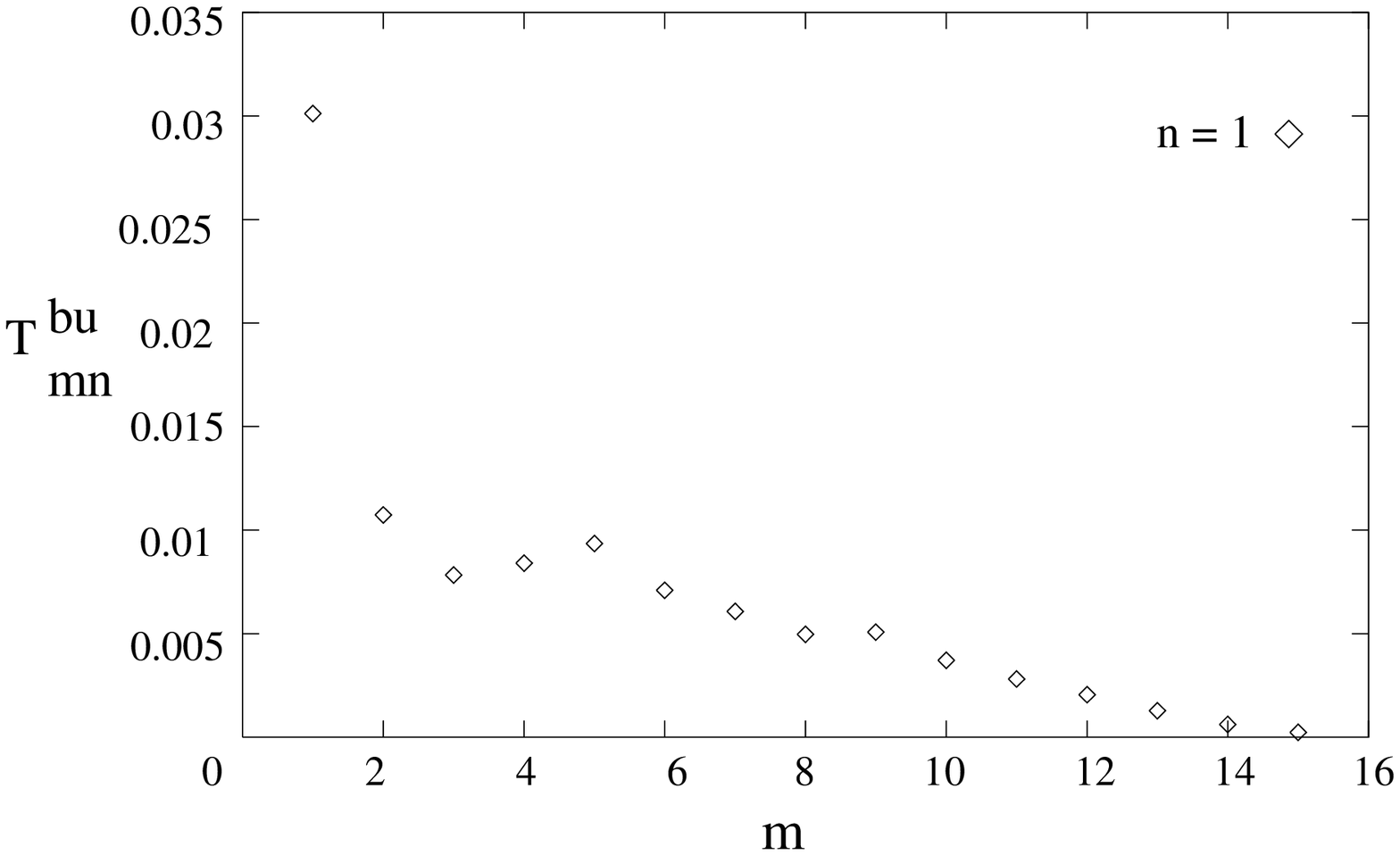,width=0.8\textwidth}}}
\caption{}
\label{fig:Tbun1}
\end{figure}

\begin{figure}[h]
\centerline{\mbox{\psfig{file=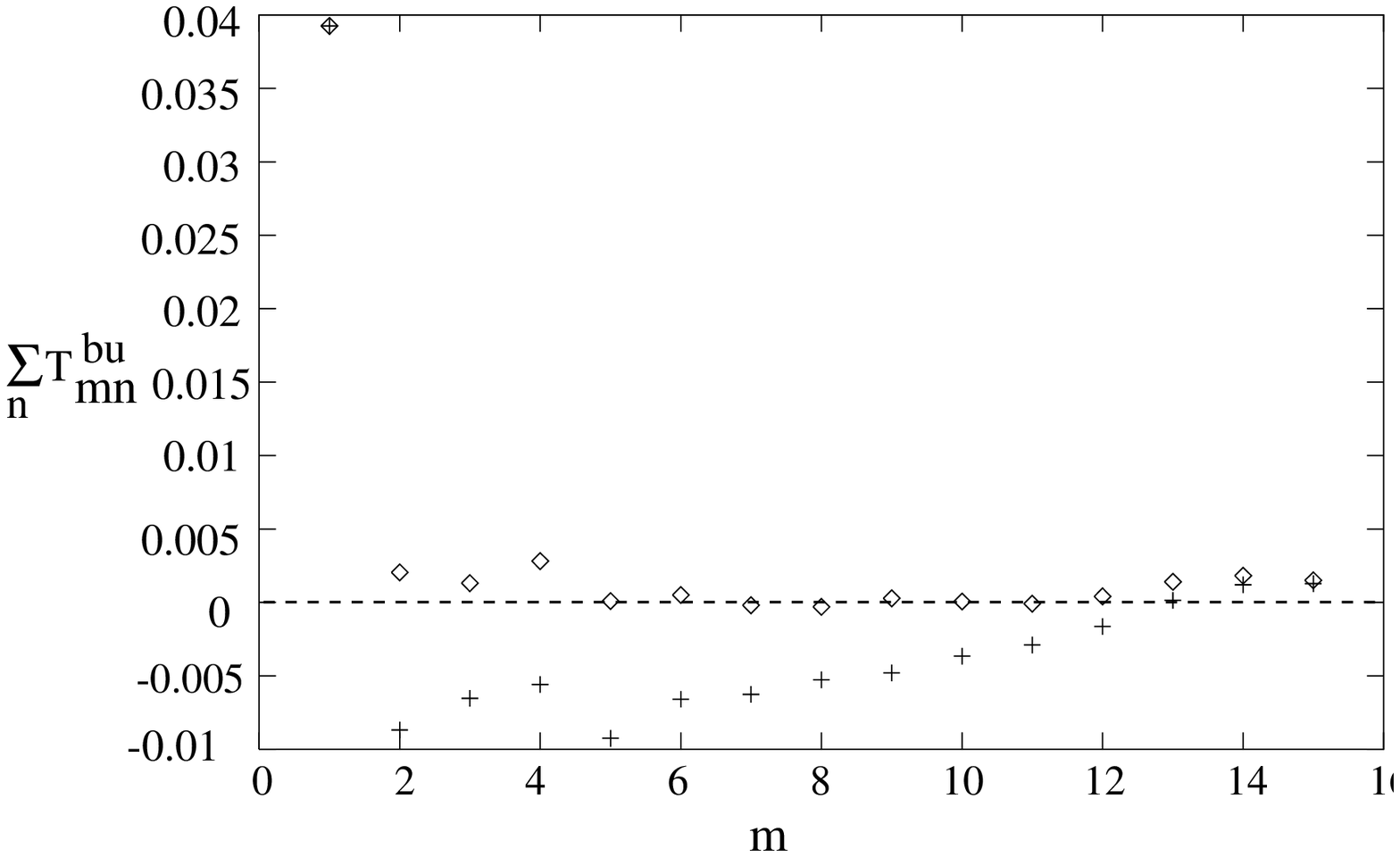,width=0.8\textwidth}}}
\caption{}
\label{fig:Tbusum}
\end{figure}

\begin{figure}[h]
\centerline{\mbox{\psfig{file=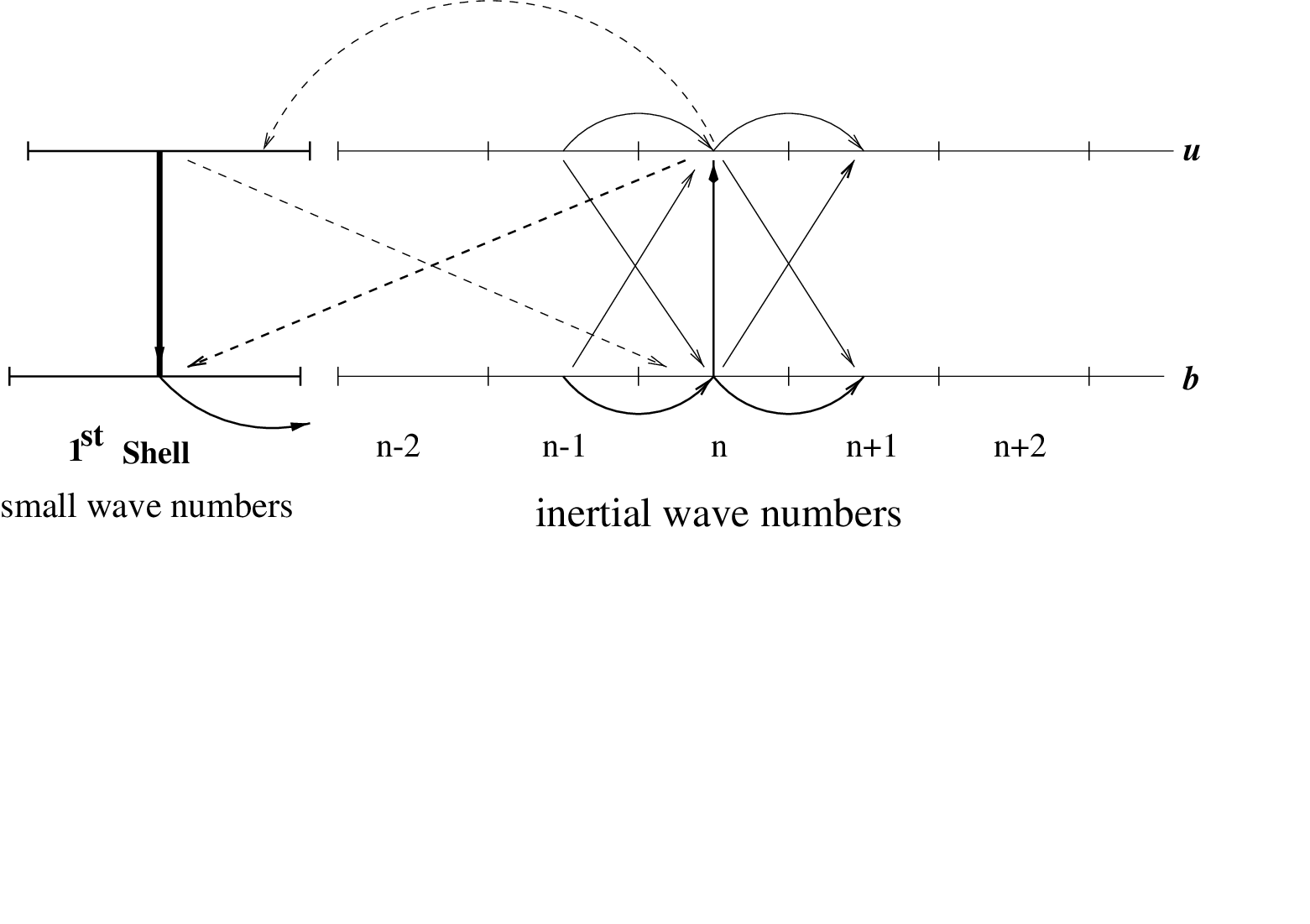,width=0.9\textwidth}}}
\caption{}
\label{fig:T}
\end{figure}

If another solution ${\cal{\slash{R}}}^{uu}({\bf k|p|q})$ differs from
${\cal{\slash{S}}}^{uu}({\bf k|p|q})$ by an arbitrary function
$X_\Delta$, i.e., ${\cal{\slash{R}}}^{uu}({\bf k|p|q})=
{\cal{\slash{S}}}^{uu}({\bf k|p|q})+X_\Delta$, then by inspection we
can easily see that $X_\Delta$ gets added to other two mode-to-mode
transfer as well.  This conclusion is depicted in Fig.~\ref{fig:S}.
As shown in the figure, $X_\Delta$ simply ``circulates'' between the
modes, hence it does not appear in the effective energy transfer. Note
that $X_\Delta$ could depend upon the wavenumber triad ({\bf k, p, q,})
and the Fourier components {\bf u(k), u(p), u(q)}.  It may be possible
to determine $X_{\Delta}$ by imposing additional constraints of
rotational invariance, Galilean invariance, finiteness, 
etc.~\cite{Dar:thesis}. We discuss in Section \ref{s:formalism} that
the circulating transfer plays no role in modal energy change, hence
we will not dwell upon the problem of its determination. In Section
\ref{s:flux_shell_to_shell} we also show that the cascade rates and
the shell-to-shell transfers can be written in terms of effective
mode-to-mode transfer rates (${\cal{\slash{S}}}^{\alpha \beta}$).

We can similarly show that (1) the ME transfer rate
${\cal{\slash{R}}}^{bb}$ from the mode {\bf b(p)} to {\bf b(k)}
mediated by mode {\bf u(q)} is given by
\be
{\cal{\slash{R}}}^{bb}={\cal{\slash{S}}}^{bb}+Y_{\Delta}, 
\ee
and
the conversion rate of KE from the mode
{\bf u(p)} to ME of mode {\bf b(k)} mediated by mode {\bf b(q)},
${\cal{\slash{R}}}^{bu}$ is given by
\be
{\cal{\slash{R}}}^{bu}={\cal{\slash{S}}}^{bu}+Z_{\Delta}.
\ee
The terms ${\cal{\slash{R}}}^{bb}$ and ${\cal{\slash{R}}}^{bu}$ are
given by Eqs.~(\ref{eqn:Sbb_mode}) and (\ref{eqn:Sbu_mode}) respectively.
The quantities $Y_{\Delta}$ and $Z_{\Delta}$ are the ``circulating
transfers''.  As shown in Fig.~\ref{fig:S}, $Y_{\Delta}$ circulates
around ${\bf b(k) \rightarrow b(p) \rightarrow b(q) \rightarrow
b(k)}$, while $Z_{\Delta}$ circulates around ${\bf u(p) \rightarrow
b(k) \rightarrow u(q) \rightarrow b(p) \rightarrow u(k) \rightarrow
b(q) \rightarrow u(p)}$.  The details of the above derivation
can be found in \cite{Dar:thesis,Dar:modetomode}.


\begin{center}
FIGURE CAPTIONS
\end{center}

\vspace{1.0cm}

\noindent 
Fig.~1:
The mode-to-mode energy transfers sought to be determined.

\vspace{1.0cm}
\noindent Fig.~2:
The energy transfer between a pair of modes in a triad can be
expressed as a sum of a circulating transfer and an effective
mode-to-mode transfer.

\vspace{1.0cm}
\noindent Fig.~3: 
Various energy fluxes defined in this section.  The arrows indicate
the directions of energy transfers corresponding to a positive
flux.  These fluxes are computed using data of numerical simulations
and are shown in Fig.~8.

\vspace{1.0cm}
\noindent Fig.~4: 
The circulating transfer and effective mode-to-mode transfer from shell
{\it n} to shell {\it m} with the arrows pointing towards the direction
of energy transfer.

\vspace{1.0cm}
\noindent Fig.~5:
Evolution of the KE and the ME for the simulations on grid sizes
$512^{2}$ and $128^{2}$.  For $512^{2}$ a quasi-steady magnetic energy
is obtained over the period of the simulation.  It is demonstrated in
the $128^{2}$ simulation that a quasi-steady magnetic energy
eventually decays.

\vspace{1.0cm}
\noindent Fig.~6:
The averaged KE spectrum $E^{u}(k)$ and 
ME spectrum $E^{b}(k)$: Averaging is done over
15 time units. 

\vspace{1.0cm}
\noindent Fig.~7:
The plots of various fluxes vs. wavenumbers.

\vspace{1.0cm}
\noindent Fig.~8:
The schematic illustration of numerically calculated values of
fluxes averaged over 15 time units.  The values shown here are
for $K=20$, a wavenumber within the inertial range.

\vspace{1.0cm}
\noindent Fig.~9:
The energy transfer rate $T^{bb}_{mn}$ from the $n^{th}$ 
{\it b}-shell to the $m^{th}$ {\it b}-shell.  The loss of energy 
from the $n^{th}$ {\it b}-shell to the $m^{th}$ {\it b}-shell is 
defined to be positive.

\vspace{1.0cm}
\noindent Fig.~10: 
The energy transfer rate $T^{uu}_{mn}$ from the
$n^{th}$ {\it u}-shell to the $m^{th}$ {\it u}-shell. 
The boxed points represent energy transfer
from the $n^{th}$ {\it u}-shell to the $1^{st}$ {\it u}-shell.

\vspace{1.0cm}
\noindent Fig.~11: 
The energy transfer rate $T^{bu}_{mn}$ from the
$n^{th}$ {\it u}-shell to the $m^{th}$ {\it b}-shell. 

\vspace{1.0cm}
\noindent Fig.~12: 
The energy transfer rate from the $1^{st}$ {\it
u}-shell to the {\it b}-shells.

\vspace{1.0cm}
\noindent Fig.~13: 
The diamonds ($\diamond$) represent the net energy
transfer into a {\it b}-shell from all the {\it u}-shells. The pluses
($+$) represent the energy transfer into a {\it b}-shell from shells
$n \ge 2$.

\vspace{1.0cm}
\noindent Fig.~14:
A schematic representation of the direction and the magnitude of
energy transfer between {\it u}-shells and {\it b}-shells. The
relative magnitudes of the different transfers have been represented by
the thickness of the arrows.  The non-local transfers with the
$1^{st}$ shell have been shown by dashed lines.

\newpage

\newpage

\newpage

\newpage

\newpage

\newpage

\newpage

\newpage

\newpage

\newpage

\newpage

\newpage

\end{document}